\documentclass[preprintnumbers,
a4paper,
superscriptaddress,
prd,nofootinbib,floatfix,12pt,
]{revtex4}
\usepackage{amsmath,amsfonts,amssymb,graphicx,natbib,bm}
\usepackage{color}
\usepackage{slashed}    
\usepackage{hyperref}
\usepackage{breakurl}
\usepackage{multirow}
\usepackage{color}
\usepackage{accents}

\begin{document}

\preprint{PITT-PACC-1620}

\title{Indirect Detection of Neutrino Portal Dark Matter}

\author{Brian Batell}
\affiliation{Pittsburgh Particle Physics, Astrophysics, and Cosmology Center, \\ 
Department of Physics and Astronomy, University of Pittsburgh, PA 15260, USA}

\author{Tao Han}
\affiliation{Pittsburgh Particle Physics, Astrophysics, and Cosmology Center, \\
Department of Physics and Astronomy, University of Pittsburgh, PA 15260, USA}

\author{Barmak Shams Es Haghi}
\affiliation{Pittsburgh Particle Physics, Astrophysics, and Cosmology Center, \\
Department of Physics and Astronomy, University of Pittsburgh, PA 15260, USA}


\begin{abstract}
We investigate the feasibility of the indirect detection of dark matter in a simple model using the neutrino portal.
The model is very economical, with right-handed neutrinos generating neutrino masses through the Type-I seesaw mechanism and simultaneously mediating interactions with dark matter. Given the small neutrino Yukawa couplings expected in a Type-I seesaw,  direct detection and accelerator probes of dark matter in this scenario are challenging. However, dark matter can efficiently annihilate to right-handed  neutrinos, which then decay via active-sterile mixing through the weak interactions, leading to a variety of indirect astronomical signatures. We derive the existing constraints on this scenario from {\it Planck} cosmic microwave background measurements, {\it Fermi} dwarf spheroidal galaxies and Galactic Center gamma-rays observations, and AMS-02 antiprotons observations, and also discuss the future prospects of {\it Fermi} and the Cherenkov Telescope Array. Thermal annihilation rates are already being probed for dark matter lighter than about 50 GeV, and this can be extended to dark matter masses of 100 GeV and beyond in the future. 
This scenario can also provide a dark matter interpretation of the {\it Fermi} Galactic Center gamma ray excess, and we confront this interpretation with other indirect constraints. Finally we discuss some of the exciting implications of extensions of the minimal model with large neutrino Yukawa couplings and Higgs portal couplings. 
\end{abstract}
\maketitle
\newpage

\baselineskip=15pt
\tableofcontents

\baselineskip=18pt

\section{Introduction}

A wide array of gravitational phenomena over a range of cosmological scales strongly supports the hypothesis of dark matter (DM)~\cite{Jungman:1995df,Bergstrom:2000pn,Bertone:2004pz}. 
There is, however, no firm evidence that DM couples to ordinary matter other than through gravity, and the search for such non-gravitational DM interactions has become one of the main drivers in particle physics today. 
Neutrinos ($\nu$) in the Standard Model (SM) may be identified as a component of DM, since they are color-singlet, electrically neutral cosmic relics. However, the smallness of the lightest neutrino mass makes them relativistic at freeze-out in the early universe, and thus incompatible with current observations to account for the majority of the cold DM. One therefore must seek a solution beyond the SM.
Since we do not know how DM couples (if at all) to the SM, it is important to explore a variety of models to understand in a comprehensive manner how non-gravitational DM interactions may manifest~\cite{Feng:2010gw}. 

Since DM is presumably electrically neutral, it may be either the neutral component of an electroweak multiplet, as in the well motivated weakly interacting massive particle (WIMP) paradigm,  or alternatively it may be a Standard Model (SM) gauge singlet state. In the latter case of gauge singlet DM, an economical and predictive mechanism for mediating DM interactions to the SM is provided by the so-called ``portals''$-$ renormalizable interactions of DM through gauge singlet SM operators. There are only three such portals in the SM$-$the Higgs portal~\cite{Silveira:1985rk,Patt:2006fw}, the vector portal~\cite{Galison:1983pa,Holdom:1985ag}, and the neutrino portal~\cite{seesaw}.  As applied to DM, the Higgs portal~\cite{Silveira:1985rk,McDonald:1993ex,Burgess:2000yq,Assamagan:2016azc}, and the vector portal~\cite{Pospelov:2007mp,ArkaniHamed:2008qn,Alexander:2016aln} have been extensively investigated, while the neutrino portal option has received comparatively little attention, despite the strong motivation  due to its connection to neutrino masses. 
In this paper we will examine a minimal model of neutrino portal DM in the simplest setup of a Type-I seesaw scenario \cite{seesaw}.

The neutrino portal to DM relies on DM interactions being mediated by the right-handed neutrinos (RHNs). Since the RHNs are responsible for generating neutrino masses, one may typically expect the DM interaction strength with the SM to be very small since it is governed by the neutrino Yukawa coupling. In this case it is challenging to probe neutrino portal DM in accelerator experiments or in direct detection experiments. On the other hand, the DM coupling to the 
RHN can be sizable, thereby facilitating the efficient annihilation of DM to pairs of RHNs. This allows DM to be produced thermally in the early universe with the observed relic abundance and furthermore presents an opportunity to test the scenario through a variety of indirect detection channels. In this work we investigate the indirect detection signatures of neutrino portal DM. 
The scenario investigated here was first proposed in Ref.~\cite{Pospelov:2007mp} and falls into the class of ``secluded'' DM scenarios.
Some aspects of the thermal cosmology were investigated in Ref.~\cite{Tang:2016sib}.
In regards to indirect detection signatures, Ref.~\cite{Tang:2015coo} explored a possible interpretation of the {\it Fermi} Galactic Center gamma ray excess~\cite{TheFermi-LAT:2015kwa,Goodenough:2009gk,Hooper:2010mq,Daylan:2014rsa,Calore:2014xka} in terms of the DM annihilation to RHNs.
Recently, Ref.~\cite{Campos:2017odj} investigated the limits from gamma ray observations on DM annihilation to RHNs, although did not explore the implications for specific particle physics models. 
Extensions of the simplest scenario, which include additional states and/or interactions have also been discussed in Refs.~\cite{Falkowski:2009yz,Kang:2010ha,Falkowski:2011xh,Cherry:2014xra,Macias:2015cna,Gonzalez-Macias:2016vxy,Escudero:2016tzx,Escudero:2016ksa,Allahverdi:2016fvl,Bertoni:2014mva,Ibarra:2016fco}. 
Our work provides a comprehensive and updated analysis of the indirect detection phenomenology of neutrino portal DM. In particular, we present constraints from {\it Planck} cosmic microwave background (CMB) measurements, {\it Fermi} dwarf spheroidal galaxies and Galactic Center gamma-rays studies, and AMS-02 antiproton observations, and also describe the future prospects for {\it Fermi} and the Cherenkov Telescope Array. Thermal relic annihilation rates are already constrained for DM masses below about 50 GeV.  This scenario can also provide a DM interpretation of the {\it Fermi} Galactic Center gamma ray excess, although we demonstrate that such an interpretation faces some tension from dSphs and antiproton constraints. We also describe extensions of this scenario beyond the minimal model, including scenarios with large Yukawa and Higgs portal couplings, and highlight the potentially rich physics implications in cosmology, direct detection, and collider experiments. 
Besides these probes, there is also the interesting possibility of a hard gamma-ray spectral feature that arises from the radiative decays of $N$, which could place complementary constraints in the region $m_\chi \sim m_N$, $m_N \lesssim 50$ GeV. We will comment on this possibility below, and we refer the reader to Ref.~\cite{Ibarra:2016fco} for a detailed study.

The outline of the paper is as follows. In Section~\ref{sec:model} we describe a minimal neutrino portal DM model, outline the expected range of couplings and masses, and discuss the cosmology. The primary analysis and results concerning the indirect detection limits and prospects are discussed in Section~\ref{sec:indirect}. In Section~\ref{sec:beyond} we describe several features and phenomenological opportunities present in non-minimal neutrino portal DM scenarios. Our conclusions are presented in Section~\ref{sec:outlook}.

\section{Neutrino Portal Dark Matter}
\label{sec:model}

The simplest construction beyond the Standard Model to account for the neutrino masses is the introduction of right-handed neutrinos (RHN). Beside the normal Dirac mass terms with the Yukawa interactions, the RHN can also have a Majorana mass term since it is a SM gauge singlet. This is the traditional Type-I  seesaw mechanism \cite{seesaw}. For the same reason of its singlet nature, $N$ can serve as a mediator to the dark sector via the neutrino portal.
A simple model of neutrino portal DM based on the Type-I seesaw 
contains three new fields, $N,\chi,\phi$, where $N$ and $\chi$ are two component Weyl fermions and $\phi$ is a real scalar field. 
They are charge-neutral with respect to the SM gauge interactions.
The fermion $N$ is identified as a RHN. We will assume that $\chi$ is lighter than $\phi$, and  they 
are charged under a $Z_2$ symmetry, which renders $\chi$ stable and a potential DM candidate. 

The Lagrangian has the following new mass terms and Yukawa interactions
\begin{eqnarray}
{\cal L}&  \supset & - \frac{1}{2}m_\phi^2 \phi^2 - \left[ \frac{1}{2} m_N N N + \frac{1}{2} m_\chi \chi \chi  +y L H N +\lambda N \phi \chi + {\rm h.c.} \right] , 
\label{eq:type1L}
\end{eqnarray}
where $L$ and $H$ are the SM SU(2)$_L$ lepton and Higgs doublets, respectively.
There are two central features of this model. First, the RHN field $N$ serves as a mediator between the dark sector fields $\chi$, $\phi$ and the SM fields, due to the couplings $\lambda$ and $y$. This mediation allows for non-gravitational signatures of the DM and a thermal DM cosmology. Second, after the Higgs obtains a vacuum expectation value, $\langle H \rangle = v/\sqrt{2}$ with $v = 246$ GeV, a small mass for the light SM-like neutrinos is generated via the seesaw mechanism:
\begin{equation}
m_\nu \sim \frac{y^2 v^2}{2 m_N}.
\label{eq:seesaw}
\end{equation}
Given the observed neutrino masses\footnote{In principle, we would need at least two right-handed states to generate the observed neutrino mass pattern. For our current interest, we will only focus on the lower-lying one $N$.}, the Yukawa coupling $y$ depends on the RHN mass, $m_N$. 
For instance, fixing $m_\nu \sim \sqrt{(\Delta m_\nu)^{\rm atm}} \sim 0.05$ eV suggsts a small neutrino Yukawa coupling of order
\begin{equation}
y \simeq 10^{-6}\,(m_N/v)^{1/2}. 
\label{eq:yukawa}
\end{equation}
As we will discuss in more detail shortly, the requirement of thermal freeze-out of the DM puts an upper bound on the DM and RHN mass 
less than $20$ TeV. Therefore, the Yukawa couplings that we will be interested in will generally be quite small. 
It will thus be extremely difficult to produce the DM at accelerators, or directly detect it through its scattering with SM particles. However, there is an opportunity to probe this type of DM via indirect detection, and this will be the primary focus of this paper. 

As alluded to already we will be interested in DM that is thermally produced in the early universe. The RHN mediator allows for the dark sector to couple to the SM thermal bath in the early universe. Then, provided that $m_\chi > m_N$ and that all of the particles are sufficiently light, say below ${\cal O}$(10 TeV), the DM can efficiently annihilate to RHNs,  
\begin{equation}
\chi \chi \rightarrow NN,
\label{eq:ann}
\end{equation} 
and achieve the correct relic abundance. The process Eq.~(\ref{eq:ann}) is governed by the coupling $\lambda$, which is a priori  a free parameter. The thermally averaged annihilation cross section is
\begin{equation}
\langle \sigma v \rangle =  \frac{\left[{\rm Re}({\lambda)}^2(m_\chi+m_N)+{\rm Im}({\lambda)}^2(m_\chi-m_N)\right]^2}{16 \pi [m_\phi^2 + m_\chi^2 - m_N^2]^2}\left( 1- \frac{m_N^2}{m_\chi^2}   \right)^{1/2}.
\label{eq:sigv}
\end{equation}
We observe that the annihilation cross section Eq.~(\ref{eq:sigv}) depends on the coupling $\lambda$ and the masses $m_\chi$, $m_N$, $m_\phi$.
However, the indirect detection signatures that we will investigate will depend in a detailed way only on the size of the annihilation cross section $\langle \sigma v \rangle$, which determines the rate, as well as the masses $m_\chi$ and $m_N$, which will affect the energy spectrum of the SM annihilation products. Thus, it will be more convenient to simply work with the three parameters $\{\langle \sigma v \rangle, m_\chi, m_N\}$. Note that for a given set of masses, one can always obtain the desired cross section by an appropriate choice of the coupling $\lambda$ through Eq.~(\ref{eq:sigv}), provided the coupling remains perturbative. We will discuss this point in detail shortly. 

We can restrict the parameter space further if we demand that the DM saturates the observed relic density. For Majorana fermion DM the observed relic abundance is obtained for~\cite{Steigman:2012nb}
\begin{equation}
\langle \sigma v \rangle_{\rm thermal} = 2.2 \times 10^{-26} \,{\rm cm}^3\ {\rm s}^{-1}. 
\label{eq:sigmav}
\end{equation}
Once we fix the annihilation cross section to saturate the observed relic abundance, then all of the physics can be characterized in terms of the two masses $m_\chi$ and $m_N$. Parameter choices that predict cross sections smaller than (\ref{eq:sigmav}) overproduce the DM.

We now discuss the expected range of masses and couplings of the new states in the model. 
A first constraint comes from demanding that the coupling $\lambda$ be perturbative and thus the theory be predictive. 
Assuming $m_N \ll m_\chi$, the partial-wave perturbative unitarity bound for the DM annihilation amplitude requires that $\lambda < \sqrt{4 \pi}$.
The over-closure and perturbative unitarity constraints lead to the bound
\begin{equation}
m_\chi \lesssim \sqrt{ \frac{\pi}{4 \langle \sigma v \rangle_{\rm thermal}}} \approx 20\ {\rm  TeV},
\end{equation}
which is in broad agreement with the general analysis of Ref.~\cite{Griest:1989wd}.
Furthermore, there are a variety of limits on the right-handed neutrinos $N$, which depend on its mass and mixing angle with active neutrinos. In particular, for seesaw motivated mixing angles, the lifetime of $N$ is typically longer than ${\cal O}(1\,{\rm s})$ for $m_N \lesssim 1$ GeV,  and is thus constrained by Big Bang Nucleosynthesis~\cite{Boyarsky:2009ix,Ruchayskiy:2012si}. Then, considering $m_\chi > m_N$ in order to obtain and efficient DM annihilation cross section we will consider in this paper masses in the range
\begin{equation}
1\ {\rm   GeV} < m_N  < m_\chi \lesssim 20\ {\rm  TeV}.
\end{equation}

The discussion above assumes a standard thermal history for the DM particle $\chi$, which relies on $\chi$ being in equilibrium with the plasma. Since the dark sector particles $\chi$ and $\phi$ have no direct couplings to the SM, it is the RHN that is ultimately responsible for keeping $\chi$ and $\phi$ in equilibrium. It is therefore important that $N$ remain in equilibrium with the SM during the freezeout process. The relevant processes to consider are the decay and inverse decays of $N$ to the SM. This question has been investigated recently in Ref.~\cite{Tang:2015coo}\footnote{See Ref.~\cite{Bandyopadhyay:2011qm} for a similar discussion in the context of right-handed sneutrino DM.}. For Yukawa couplings dictated by the naive seesaw relation, these process are very efficient when $m_N \gtrsim m_W$, since $N$ decays through a two body process. However, if $N$ is light, $m_N \lesssim m_W$, the three body decays of $N$ become inefficient and $N$ can fall out of equilibrium. As a consequence, this fact requires an annihilation cross section that is larger than the canonical thermal relic value by some order one factor in the early universe to efficiently deplete the $\chi$ abundance, as explored in detail in Ref.~\cite{Tang:2015coo}. A detailed investigation of the cosmology is beyond the scope of this paper, but we will take the standard thermal value for the annihilation cross section as a motivated benchmark. 

Besides the terms in Eq.~(\ref{eq:type1L}), an additional Higgs portal coupling, $\phi^2 |H|^2$ is allowed in the model. This interaction provides an alternative means to keep $\phi$, $\chi$, and $N$ in thermal equilibrium with the SM.
We will assume for now that this coupling is small so that the phenomenology is dictated by the minimal neutrino portal interaction. However, a large Higgs portal coupling can lead to a variety of interesting effects, and we will discuss this topic in Section~\ref{sec:beyond}.

\section{Indirect Detection Constraints and Prospects}
\label{sec:indirect}

We now come to the main subject of this work: the constraints and prospects for indirect detection of neutrino portal DM. 
We will investigate several indirect signatures of DM annihilation in this scenario, including observations of the CMB, gamma rays, and antiprotons.
For each of these indirect probes the relevant underlying reaction 
is DM annihilation to RHNs as in Eq.~(\ref{eq:ann}), followed by the weak decays of the RHNs to SM particles due to mixing. We will thererfore require the energy spectrum $dN/dE$ per DM annihilation in the photon, electron and antiproton channels as an input to our further analysis below. 
To compute these spectra we first simulate the decay of RHNs to SM particles in the $N$-rest frame using \texttt{MadGraph5\_aMC@NLO}~\cite{Alwall:2014hca} in conjunction with the \texttt{SM\_HeavyN\_NLO} model files~\cite{Alva:2014gxa, Degrande:2016aje}. 
These partonic events are then passed to~\texttt{Pythia 8}~\cite{Sjostrand:2007gs} for showering and hadronization, thereby yielding the prediction for the resulting photon, electron, and antiproton spectrum coming from the $N$ decay, $dN'_i/dE'$ for $i = \gamma, e^-, \bar p$. These events are then boosted to the DM rest frame according to the formula (see, {\it e.g.}, Refs.~\cite{Mardon:2009rc,Agrawal:2014oha,Elor:2015tva,Elor:2015bho} for the case of massless particles):
\begin{eqnarray}
\frac{dN_i}{dE} =  \int_{\gamma(E-\beta\sqrt{E^2-m^2})} ^{\gamma(E+\beta\sqrt{E^2-m^2})} \frac{dE'}{2\beta\gamma \sqrt{E'^2-m^2}}\,\frac{dN'_i}{dE'}, ~~~~~~ \gamma = (1-\beta^2)^{-1/2} = m_\chi/m_N,
\label{eq:spectrum-boosted}
\end{eqnarray}
where $m$ is the mass of the boosted particle, i.e., photons, or electrons, antiprotons; see Appendix A for a derivation of Eq.~(\ref{eq:spectrum-boosted}).  This gives the prediction for the required spectrum in each channel. We note that spin correlations are not accounted for in our simulation, but these are expected to have only a modest effect on the broad continuum spectra of interest to us (see Ref.~\cite{Elor:2015tva} for an explicit example where this expectation is borne out).

\begin{figure}[t]
\centering
\includegraphics[width=0.45\textwidth ]{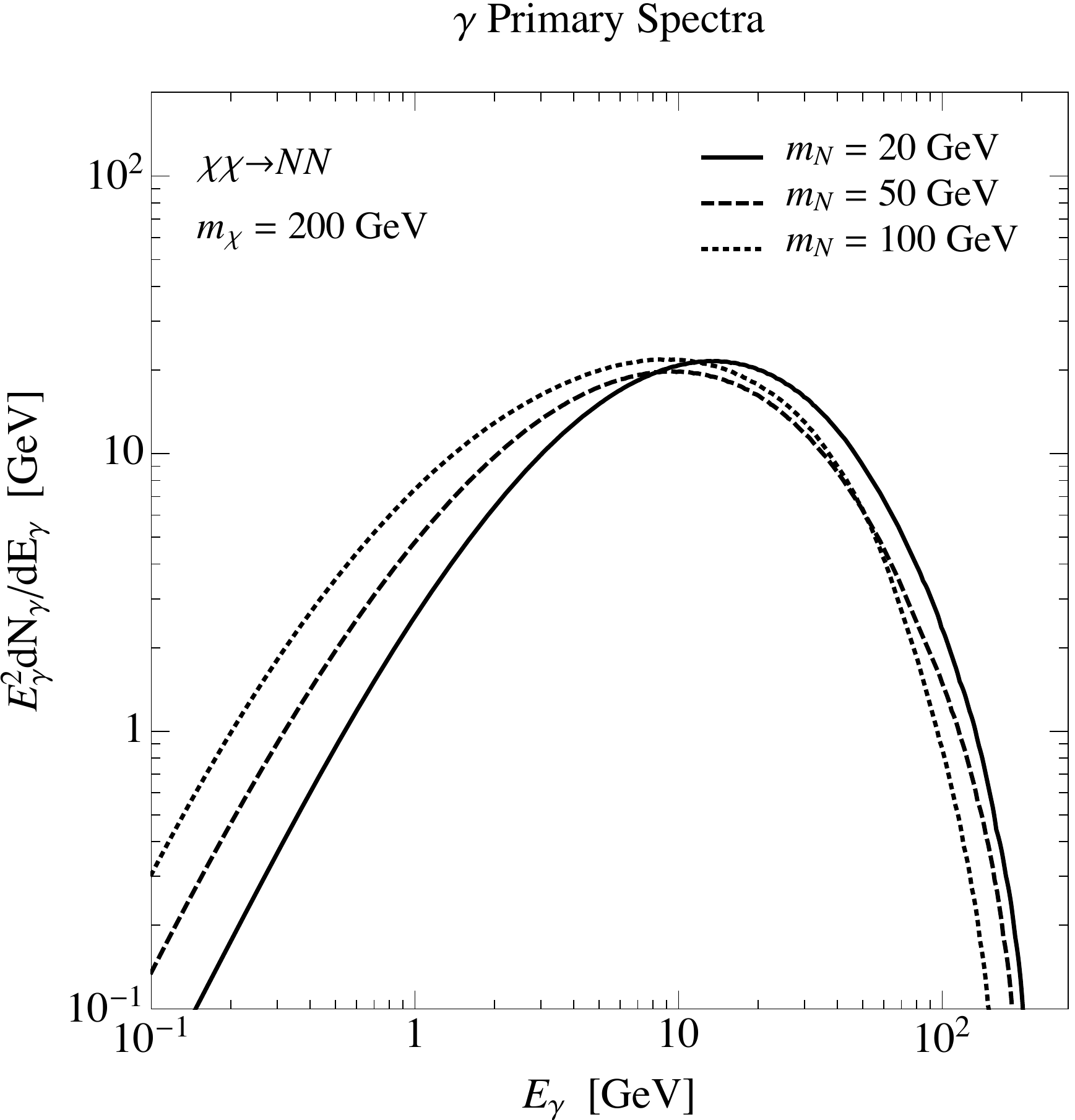}~
\includegraphics[width=0.45 \textwidth ]{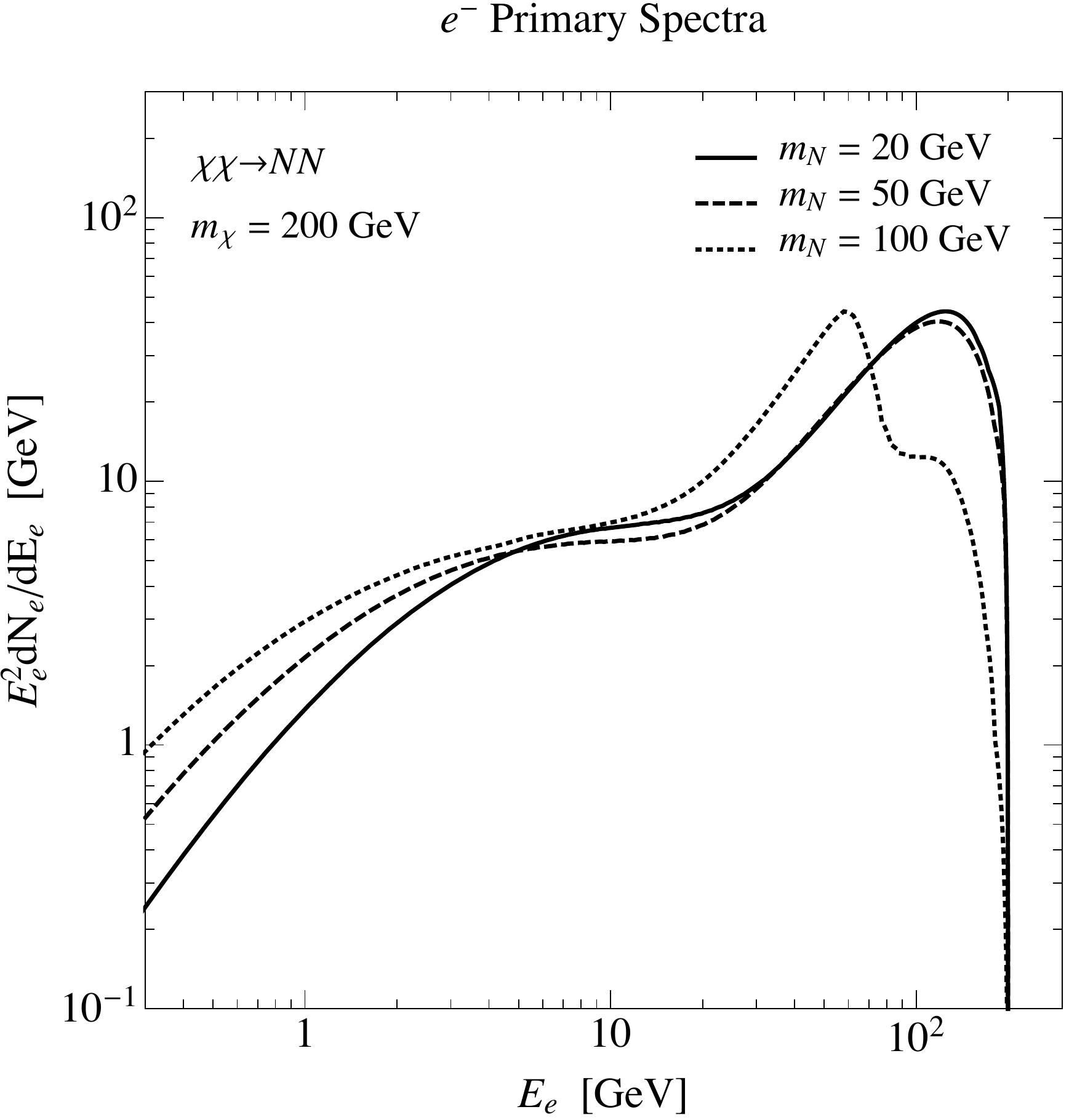} \\
\includegraphics[width=0.45 \textwidth ]{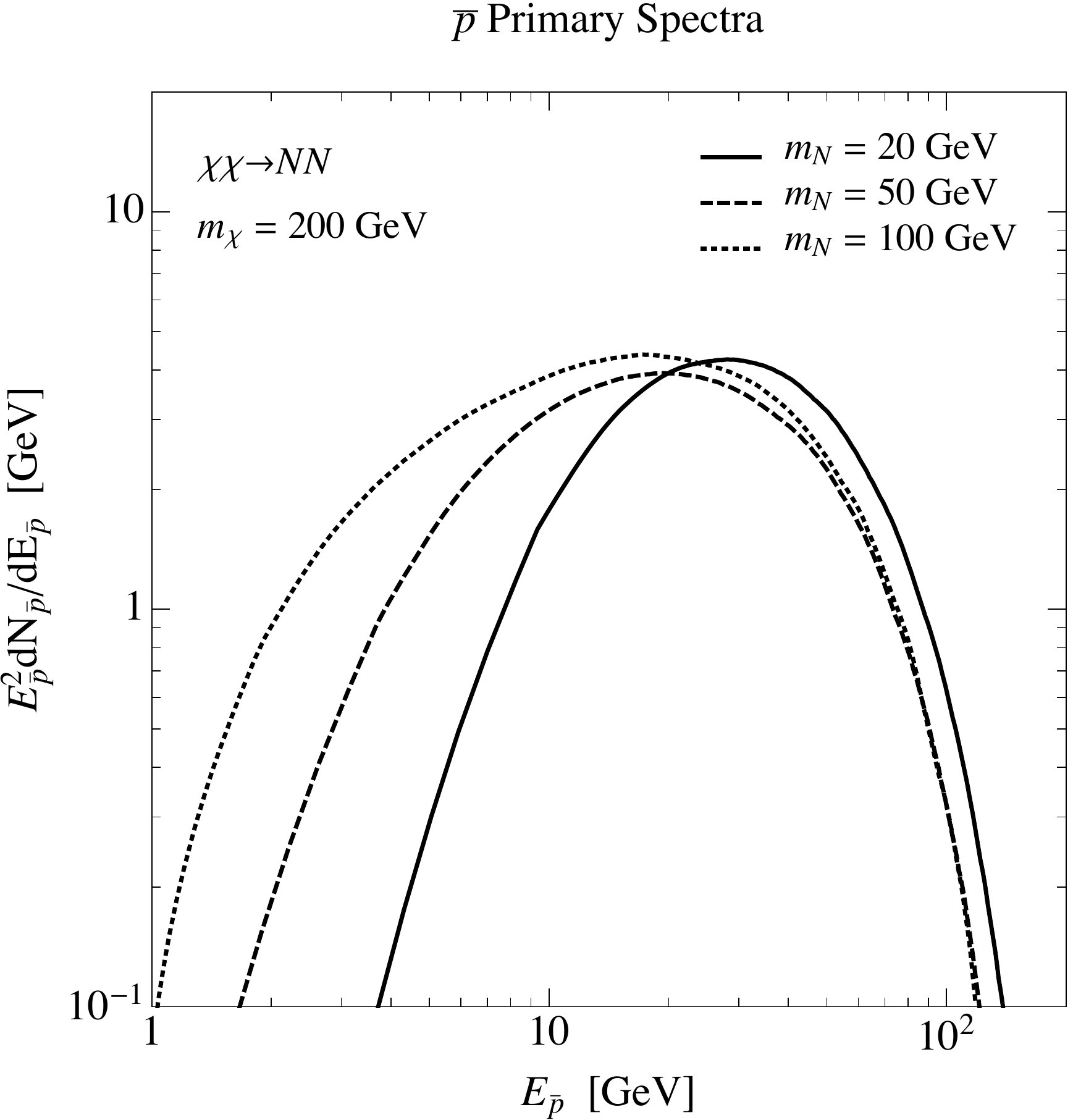} \\
\caption{\footnotesize 
Continuum $\gamma$-ray, electron, and antiproton spectra $E_i^2 dN_i /dE_i$ versus $E_i$ ($i = \gamma, e^-, \bar p$) for DM mass $m_\chi = 200$ GeV and RHN masses $m_N = 20$ GeV (solid), 50 GeV (dashed), 100 GeV (dotted). The RHN is assumed to couple to the electron-type lepton doublet. }
\label{fig:spectra}
\end{figure}

We display in Figure~\ref{fig:spectra} examples of the predicted continuum $\gamma$-ray, electron, and antiproton spectra for ($E_i^2 dN_i /dE_i$ versus $E_i$ for $i = \gamma, e^-, \bar p$), where we have fixed the DM mass to be $m_\chi = 200$ GeV and chosen three values for the RHN masses $m_N = 20$ GeV (solid), 50 GeV (dashed), 100 GeV (dotted). Here we have assumed that $N$ couples solely to the first generation (electron-type) lepton doublet. In the case of the $\gamma$-ray and antiproton spectrum, one observes a broad spectrum that peaks in the ${\cal O}(10~{\rm GeV})$ range. The location of the peak is largely dictated by the DM mass, which controls the total injected energy. There is a mild sensitivity to the RHN mass, with harder spectra resulting from a larger mass gap between the DM and RHN. For the electron case, in addition to the continuum component, there is a hard component resulting from the primary $N \rightarrow W e$ decay, which is clearly seen in Figure~\ref{fig:spectra}.

In this work we will restrict to the case in which $N$ couples to the electron-type lepton doublet, but it is worth commenting on the cases of couplings to muon and/or tau flavor.
In these cases, we have checked that the continuum spectra is very similar to the electron-flavor case, as is expected since these particles dominantly originate from decay of the electroweak bosons. 
The primary difference for muon or tau-flavor couplings will be the absence of the hard electron component from the primary $N$ decay. The electron spectrum will be used below as an input to the CMB bounds, so one may expect a mild difference in the resulting limits in the case of muon or tau flavor couplings.

We now present in turn the constraints on neutrino portal DM from the {\it Planck} cosmic microwave background measurements, {\it Fermi} observations of gamma rays from the Galactic Center and from dwarf spheroidal galaxies, and AMS-02 observations of antiprotons. A summary of these constraints, as well as a discussion of other indirect searches not considered here, and an analysis of the future prospects, is presented below in Section~\ref{sec:summary}.

\subsection{Cosmic Microwave Background}
\label{sec:CMB}

The Cosmic Microwave Background (CMB) provides a sensitive probe of DM annihilation around the epoch of recombination. In particular, if the annihilation products include energetic electrons and photons, the photon-baryon plasma can undergo significant heating and ionization as these particles are injected into the bath, modifying the ionization history and altering the temperature and polarization anisotropies. Using precise measurements of the CMB by a number of experiments, including 
WMAP~\cite{Hinshaw:2012aka}, SPT~\cite{Story:2012wx,Hou:2012xq}, ACT~\cite{Sievers:2013ica}, and {\it Planck}~\cite{Ade:2015xua}, robust, model-independent constraints on DM annihilation have been derived by several groups~\cite{Padmanabhan:2005es,Zhang:2006fr,Galli:2009zc,Slatyer:2009yq,Kanzaki:2009hf,Hisano:2011dc,Hutsi:2011vx,Galli:2011rz,Finkbeiner:2011dx,Slatyer:2012yq,Galli:2013dna,Lopez-Honorez:2013lcm,Madhavacheril:2013cna,Slatyer:2015jla,Slatyer:2015kla}. 

The relevant quantity of interest for DM annihilation during recombination is the energy absorbed by the plasma per unit volume per unit time at redshift $z$, 
\begin{equation}
\frac{dE}{dV dt} = \rho_{\rm c}^2\, \Omega_\chi^2\, (1+z)^6 \left[f(z) \frac{\langle \sigma v \rangle}{m_\chi} \right],
\label{eq:CMB}
\end{equation}
where $\rho_{\rm c}$ is the critical density of the Universe today and $\Omega_\chi$ is the DM density parameter today. Production of neutrinos as daughter particles and free-streaming of electrons and photons
after creation until their energy is completely deposited into the intergalactic medium (IGM) (via photoionization, Coulomb scattering, Compton processes, bremsstrahlung and recombination) affect the the efficiency of energy deposition. This is accounted for in Eq.~(\ref{eq:CMB}) by the efficiency factor, $f(z)$, which gives the fraction of the injected energy that is deposited into the IGM at redshift $z$ and depends on the spectrum of photons and electrons arising from DM annihilations. Furthermore, since the CMB data are sensitive to energy injection over a narrow range of redshift, {\it i.e.}, $1000-600$, $f(z)$ can be well-approximated by a constant parameter $f_{\text{eff}}$. 

The additional energy injection from DM annihilation in Eq.~(\ref{eq:CMB}) alters the free electron fraction (the abundance ratio of free electrons to hydrogen), which in turn affects the ionization history. These effects are quantitatively accounted for with new terms in the Boltzmann equation describing the evolution of the free electron fraction. The additional terms are added to the baseline $\Lambda$CDM code and used to derive limits on the energy release from DM annihilation. 
{\it Planck} sets a limit on the particle physics factors in Eq.~(\ref{eq:CMB})
\begin{eqnarray}
f_{\text{eff}}(m_{\chi})\frac{\langle\sigma v\rangle}{m_\chi} < 4.1 \times10^{-28}\, \text{cm}^3 \,\text{s}^{-1}\, \text{GeV}^{-1},
\label{eq:planck}
\end{eqnarray}
which is obtained from temperature and polarization data (TT,TE,EE+lowP) \cite{Ade:2015xua}. 

\begin{figure}[t]
\centering
\includegraphics[width=0.55 \textwidth ]{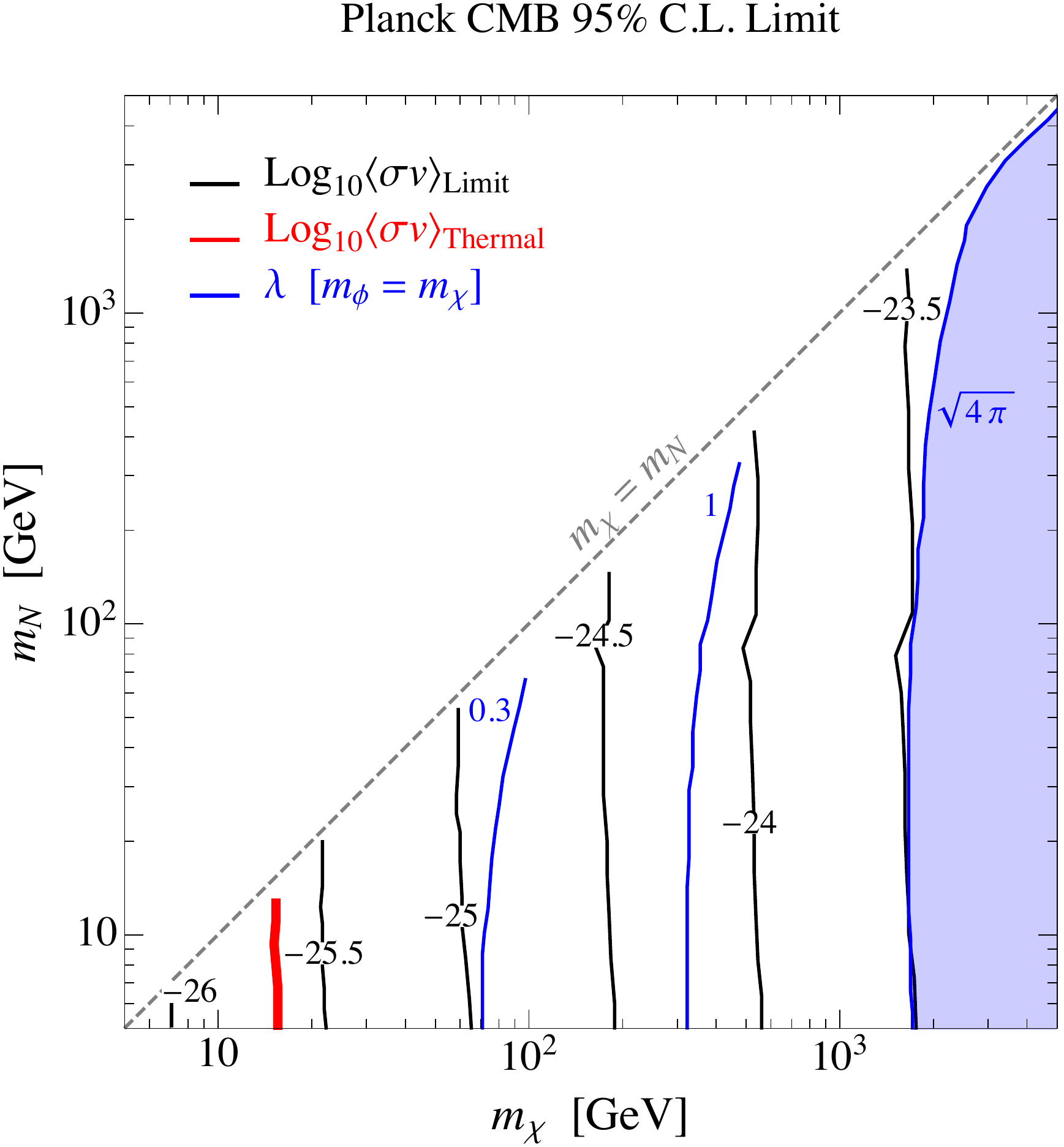}
\caption{\footnotesize 
Contours of the $95\%$ C.L. upper limit on $\log_{10}\left[\langle \sigma v \rangle/ ({\rm cm}^3\, {\rm s^{-1}})\right]$ in the $m_\chi - m_N$ plane (black curves) from {\it Planck}~\cite{Ade:2015xua}. 
The thick (red) line indicates the region where the cross section limit is equal to the thermal relic value of Eq.~(\ref{eq:sigmav}). The vertical (blue) lines show the limits on the minimum value of the coupling constant $\lambda$. The shaded (blue) region indicates the perturbative unitarity bound. 
}
\label{fig:CMBcontour}
\end{figure}

To apply the {\it Planck} constraints of Eq.~(\ref{eq:planck}) to the neutrino portal DM model, it remains to compute the efficiency factor $f_{\text{eff}}(m_\chi)$ in our model.  We use the results of Ref.~\cite{Slatyer:2015jla}, which provides $f^{\gamma (e^{-})}_{\text{eff}}(E)$ curves for photons and electrons to compute a weighted average with the photon/electron spectrum $(dN/dE)_{\gamma, e^-}$ predicted in our model according to
\begin{eqnarray}
f_{\text{eff}}(m_\chi)=\frac{1}{2 m_\chi} \displaystyle{\int_{0}^{m_\chi}\!\! dE\, E \left[2f_{\text{eff}}^{e^-}(E)\left(\frac{dN}{dE}\right)_{e^-} + f_{\text{eff}}^{\gamma}(E)\left(\frac{dN}{dE}\right)_\gamma\right]}.
\label{eq:feff}
\end{eqnarray}
\break
The photon and electron spectra for each DM and RHN mass point are computed with Monte Carlo simulation described at the beginning of this section and are displayed for a few benchmarks in Figure~\ref{fig:spectra}.
Using these spectra and Eqs.~(\ref{eq:planck}) and (\ref{eq:feff}), we obtain a limit on the annihilation cross section from the CMB as a function of $m_\chi$ and $m_N$. These limits are displayed in Figure~\ref{fig:CMBcontour}
as contours of the $95\%$ C.L. upper limit on $\log_{10}\left[\langle \sigma v \rangle/ ({\rm cm}^3\, {\rm s^{-1}})\right]$ (black curves) from the CMB from {\it Planck}~\cite{Ade:2015xua} in the $m_\chi - m_N$ plane. 
The thick (red) line indicates the region where the cross section limit is equal to the thermal relic value of Eq.~(\ref{eq:sigmav}). The constraints on the annihilation cross section are translated to 
limits on the minimum value of the coupling constant $\lambda$ (which occurs for $m_\phi = m_\chi$) as shown by the vertical (blue) lines. The shaded (blue) region indicates where the perturbative unitarity bound is violated, $\lambda > \sqrt{4\pi}$.
Since the efficiency factor $f_{\text{eff}}$ is essentially constant over a broad range of $m_\chi$, Eq.~(\ref{eq:planck}) implies that the limit on $\langle \sigma v\rangle$ scales with $m_\chi$ irrespective of the value of $m_N$, and this feature is clearly present in Figure~\ref{fig:CMBcontour}. We observe that {\it Planck} is able to constrain the thermal relic value based on Eq.~(\ref{eq:sigmav}) for DM masses below about 20 GeV. 
A small feature in the limit contour is apparent in the region near $m_W \lesssim m_N \lesssim m_Z$. This is a consequence of the dominance of the two body decay to $N \rightarrow W \ell$ in this small mass window. 

\subsection{Gamma rays from the galactic center}
\label{sec:GC}
One of the primary signatures of DM annihilation are high-energy gamma rays. In comparison to other cosmic ray signatures involving electrically charged particles, gamma rays are essentially unperturbed by magnetic fields and the astrophysical environment as they travel to us from their source, yielding information about both the energy and location of the underlying DM reaction. One can search for both gamma ray line signatures as well as a continuum signal. While a line signature is unfortunately not present in the neutrino portal DM model, there can be a distinct continuum gamma ray signal, and this will be the subject of investigation here. 
Significant advances in our study of the gamma-ray sky have been achieved over the past several years by the {\it Fermi} Gamma Ray Space Telescope, and data from the {\it Fermi} collaboration can be used to probe DM annihilation over a wide range of models and DM masses. 
In this section we will consider gamma ray signatures from the center of the Milky Way. The Galactic Center has long been recognized as the brightest source of DM induced gamma rays, a consequence of its proximity and the rising DM density in this region. At the same time extracting a signal from this region is challenging  due to significant and not well-understood astrophysical backgrounds. 
Below we will also investigate gamma ray signals from dwarf spheroidal galaxies, which provide a cleaner, albeit dimmer, source of gamma rays. 

The quantity of interest for gamma ray signals of DM annihilation is the gamma ray flux per unit energy per unit solid angle in a given direction, $\Phi_{\gamma}(E,\hat n)$, 
where $E$ is the energy and $\hat n$ is a unit
vector along the path of the line of sight. 
The gamma ray flux can be written as 
\begin{eqnarray}
\Phi_{\gamma}(E,\hat n)=\frac{1}{4\pi} \left[ \frac{\langle\sigma v\rangle}{2m_\chi^2}\frac{dN_\gamma}{dE} \right] J(\hat n).
\label{eq:photonflux}
\end{eqnarray}
The term in square brackets in Eq.~(\ref{eq:photonflux}) above depends only on the underlying particle physics properties of the DM model, including $m_\chi$, $\langle \sigma v \rangle$, and the spectrum of photons emitted per DM annihilation $dN_\gamma/dE$. This spectrum is shown in Figure~\ref{fig:spectra} for the channel $\chi \chi \rightarrow NN$ for several choices of  $\chi$ and $N$ masses.

The quantity $J(\hat n)$ in  Eq.~(\ref{eq:photonflux}), also called the $J$-factor, depends only on astrophysics and involves an integral over the DM density profile $\rho_\chi(\bf r)$ that runs along the path of the line of sight defined by $\hat n$:
\begin{eqnarray}
J(\hat n)=\int_{\rm  l.o.s.}\rho_{\chi}^2(\textbf{r})\,dl.
\label{eq:J}
\end{eqnarray}
In practice, the $J$-factor is averaged over a particular region of interest relevant for the analysis.  
The $J$-factor depends sensitively on the DM distribution and can vary by several orders of magnitude depending on this assumption, which translates into a substantial uncertainty in the derived annihilation cross section limit. At present, there is no consensus on the expected DM halo profile. Cuspy profiles such as NFW~\cite{Navarro:1995iw,Navarro:1996gj} or Einasto~\cite{Einasto} find support from $N$-body simulations~\cite{Navarro:2008kc,Springel:2008cc}. 
These simulations only involve DM, and the inclusion of baryonic processes may significantly impact the shape of the profile, especially towards the inner region of the Milky Way. 
However, even the qualitative nature of the resulting DM distribution is a matter of debate, and it is possible that the resulting profile is either steepened~\cite{Blumenthal:1985qy,Ryden:1987ska,Gnedin:2004cx,Gnedin:2011uj} or flattened~\cite{Governato:2012fa} due to baryonic effects. Besides the assumption of the DM distribution, a separate, smaller ${\cal O}(1)$ uncertainty arises from the overall normalization of the profile, which is fixed to match the local DM density $\rho_0$~\cite{Iocco:2011jz}. 

The current situation regarding the observed gamma ray flux from the Galactic Center is somewhat murky. A number of analyses, starting from the works of Goodenough and Hooper~\cite{Goodenough:2009gk,Hooper:2010mq} and culminating most recently in the {\it Fermi} analysis~\cite{TheFermi-LAT:2015kwa}, have found a broad excess of gamma rays from the Galactic Center, which peaks in the $1-3$ GeV range. All analyses conclude that there is a highly statistically significant excess above the currently accepted diffuse background models (see for example Refs.~\cite{Daylan:2014rsa,Calore:2014xka}). However, the origin of these gamma rays is still not clear. While there has been a significant effort devoted to possible DM interpretations, recently it has been argued that the excess is more likely to be a new population of unresolved point sources, which would disfavor the simplest DM interpretations~\cite{Lee:2014mza,Bartels:2015aea,Lee:2015fea,McDermott:2015ydv} (see however \cite{Horiuchi:2016zwu}). 
It is certainly interesting to speculate on a possible DM origin, and we will carry out this exercise below in Section~\ref{sec:excess}. Here we will instead take a conservative approach and use the {\it Fermi} data to place limits on DM annihilation. 

To obtain limits on the neutrino portal DM scenario, we use the model independent results of Ref.~\cite{Hooper:2012sr}. 
In that work, four years of data from the {\it Fermi} Large Area Telescope was used to construct maps of the gamma ray flux in the region around the Galactic Center in four energy bins in the range from 300 MeV$-100$ GeV. Backgrounds templates from known point sources and emission from the Galactic Disk are then subtracted to yield the residual flux. Assuming that DM annihilation accounts for the remaining emission, the authors then place limits on DM annihilation for several choices of halo profiles. This procedure yields conservative limits since it is expected that additional background sources, such as the central supermassive black hole, unresolved point sources, and cosmic ray interactions with the gas, also contribute significantly to the residual emission. Limits on the the particle physics factor that governs the gamma ray flux, $(\langle \sigma v\rangle/m_{\chi}^2) \, \int dE \, dN_\gamma/dE$, are provided in Ref.~\cite{Hooper:2012sr}.

\begin{figure}[t]
\centering
\includegraphics[width=0.55 \textwidth ]{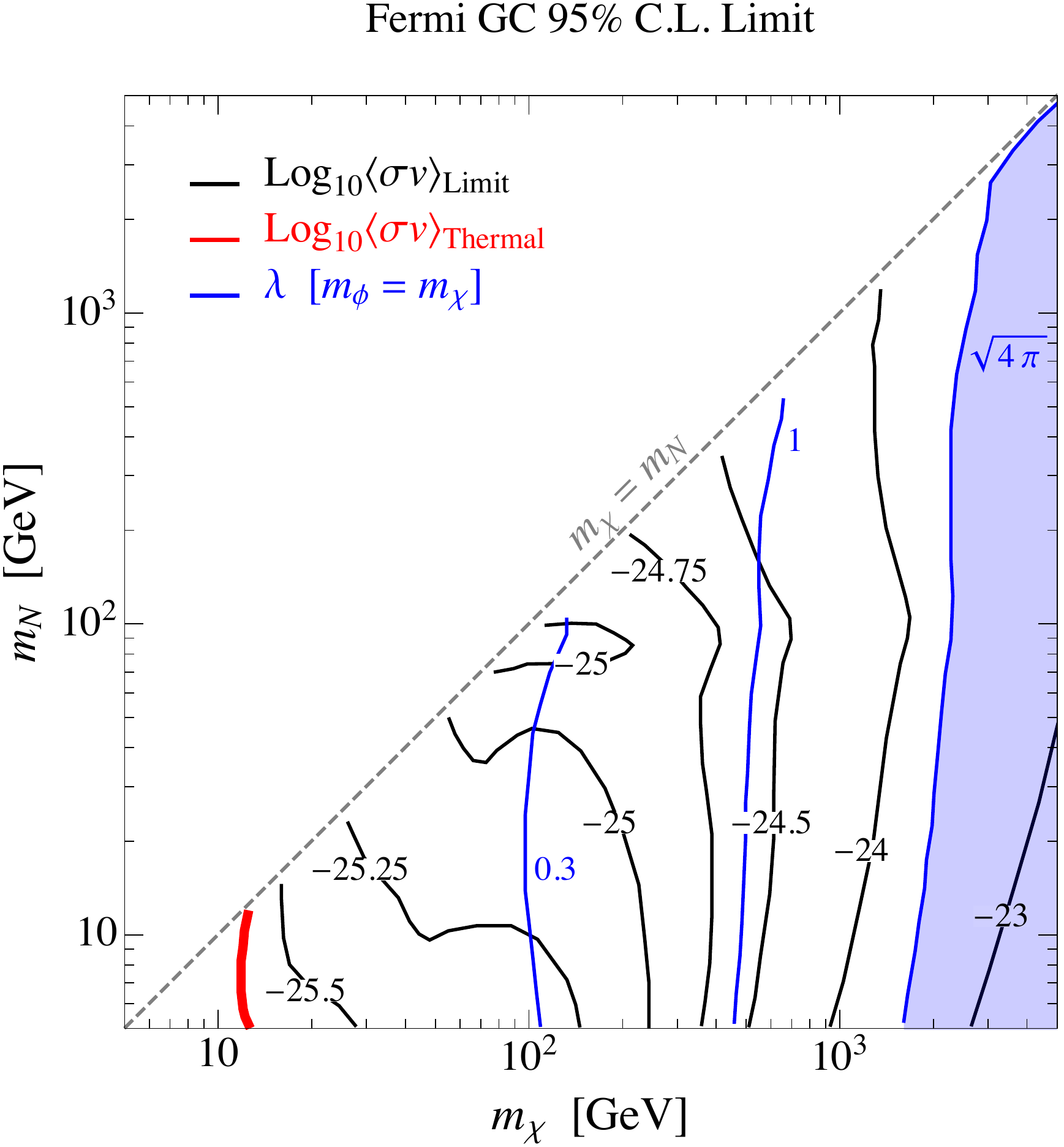}
\caption{ \footnotesize 
Contours of the $95\%$ C.L. upper limit on $\log_{10}\left[\langle \sigma v \rangle/ ({\rm cm}^3\, {\rm s^{-1}})\right]$  in the $m_\chi - m_N$ plane (black curves) from {\it Fermi} observations of gamma-rays from the Galactic Center, using the model independent results of Ref.~\cite{Hooper:2012sr}. 
The thick (red) line indicates the region where the cross section limit is equal to the thermal relic value of Eq.~(\ref{eq:sigmav}). The vertical (blue) lines show the limits on the minimum value of the coupling constant $\lambda$. The shaded (blue) region indicates the perturbative  unitarity bound.
}
\label{fig:GCRcontour}
\end{figure}

For the neutrino portal DM model, we can use these results to derive a limit on the annihilation cross section for the process $\chi\chi \rightarrow NN$ as a function of the DM and RHN mass. In Figure~\ref{fig:GCRcontour} we show contours of the 95$\%$ C.L. upper limit on the annihilation cross section in the $m_\chi - m_N$ plane labelled by the black curves. These limits are derived under the assumption of an NFW profile and local DM density $\rho_0 = 0.3$ GeV cm$^{-3}$.  We see that under these assumptions, the {\it Fermi} data probes the thermal relic cross sections of Eq.~(\ref{eq:sigmav}) for $m_\chi \lesssim$ 10 GeV (thick red contour). The constraints on the annihilation cross section are again translated to 
limits on the minimum value of the coupling constant $\lambda$ as shown by the vertical (blue) lines. The shaded (blue) region indicates the perturbative unitarity bound. 
However, we again emphasize that there are significant uncertainties associated with halo profile, and the limits will become stronger (weaker) by a factor of a few to 10 (depending of course on the detailed shape) if one assumes a contracted (cored) DM distribution~\cite{Hooper:2012sr}.  We observe a small feature near $m_W \lesssim m_N \lesssim m_Z$ where the two body decay $N \rightarrow W \ell$ dominates.

\subsection{Gamma rays from dwarf spheroidal galaxies}

Gamma ray observations of dwarf spheroidal satellite galaxies (dSphs) of the Milky Way offer a promising and complementary indirect probe of DM annihilation. There are several reasons to consider dSphs. They are DM-dominated, having mass to light ratios in the 10-2000 range. Being satellites of the Milky Way, the dSphs are nearby. There are many of them, ${\cal O}(40)$, allowing for a joint analysis to increase statistics. And, crucially, while the Galactic Center provides a significantly brighter source of DM, the dSphs are known to have substantially smaller astrophysical gamma-ray backgrounds in comparison to the Galactic Center, making them very clean sources for indirect searches.
The $Fermi$-LAT collaboration has analyzed 6 years of gamma ray data from Milky Way dSphs, finding no significant excess above the astrophysical backgrounds~\cite{Ackermann:2015zua}. Here we will discuss the implications of these null results for the neutrino portal DM scenario. 

The $Fermi$ analysis \cite{Ackermann:2015zua} is based on a joint maximum likelihood analysis of 15 dSphs for gamma ray energies in the 500 MeV - 500 GeV range.
The quantity of interest in the likelihood analysis is the energy flux,
\begin{eqnarray}
\varphi_{k,j}=\int_{E_{j,\text{min}}}^{E_{j,\text{max}}} E\,\Phi_{\gamma,k}(E)\,dE,
\label{eq:energyflux}
\end{eqnarray}
for $k$th dwarf and $j$th energy bin. For each dwarf and energy bin, $Fermi$ provides the likelihood, ${\cal L}_{k,j}$ as a function of $\varphi_{k,j}$. 
The likelihood function accounts for instrument performance, the observed counts, exposure, and background fluxes. 
For a given DM annihilation channel, the energy flux depends on $m_\chi$, $\langle \sigma v\rangle$, and $J_k$ (the $J$-factor of the dSph -- see Eq.~(\ref{eq:J})) according to Eqs.~(\ref{eq:photonflux},\ref{eq:J},\ref{eq:energyflux}), {\it i.e.}, $\varphi_{k,j}=\varphi_{k,j}(m_\chi, \langle \sigma v\rangle, J_k)$.
The likelihood for a given dwarf, ${\cal L}_{k}$, is
\begin{eqnarray}
{\cal L}_{k}(m_\chi, \langle \sigma v\rangle, J_k)={\cal L}{\cal N}(J_k\rvert \bar J_k,\sigma_k)\prod_j {\cal L}_{k,j}(\varphi_{k,j}(m_\chi, \langle \sigma v\rangle, J_k)),
\label{eq:dwarflike}
\end{eqnarray}
where ${\cal L}{\cal N}$ accounts for statistical uncertainty in the $J$-factor determination (from the stellar kinematics in the dSphs), incorporated as a nuisance parameter in the likelihood. The $Fermi$-LAT collaboration employs a log-normal distribution parameterized by $\bar J_k,\sigma_k$ :
\begin{eqnarray}
{\cal L}{\cal N}(J_k\rvert \bar J_k,\sigma_k)=\frac{1}{\text{ln}(10)J_k\sqrt{2\pi}\sigma_k}e^{-(\text{log}_{10}(J_k)-{\text{log}_{10}(\bar{J_k})})^2/2\sigma_k^2},
\label{eq:jlike}
\end{eqnarray}
where $J_k$ is the true value of the $J$-factor and $\bar {J_k}$ is the measured $J$-factor with error $\sigma_k$ on the quantity $\log_{10}{\bar {J_k}}$.
The combined likelihood for all the dwarfs is then
\begin{eqnarray}
{\cal L}(m_\chi, \langle \sigma v\rangle, \{J_i\})=\prod_k {\cal L}_{k}(m_\chi, \langle \sigma v\rangle, J_k),
\label{eq:combined}
\end{eqnarray}
where $\{J_i\}$ is the set of J-factors.

Given that no significant excess is observed, a delta-log-likelihood method is used to set limits on DM model parameters, treating the $J$-factors as nuisance parameters. 
The delta-log-likelihood $\Delta \ln{\cal L}$  is given by
\begin{eqnarray}
\Delta\ln{\cal L}(m_\chi, \langle \sigma v\rangle)=\text {ln}\,{\cal L}(m_\chi, \langle \sigma v\rangle, \{\hat{\vphantom{\rule{1pt}{10pt}}\smash{\hat{J}}}_i\})-\text {ln}\,{\cal L}(m_\chi, \widehat{\langle \sigma v\rangle},\{\Hat{J}_i\})
\label{eq:deltalike}
\end{eqnarray}
where $\widehat{\langle \sigma v\rangle}$ and $\{\Hat{J}_i\}$ are the values of $\langle \sigma v\rangle$ and $\{J_i\}$ that jointly maximize the likelihood at the given $m_\chi$, and 
$\{\hat{\vphantom{\rule{1pt}{10pt}}\smash{\hat{J}}}_i\}=\{\hat{\vphantom{\rule{1pt}{10pt}}\smash{\hat{J}}}_i(m_\chi,\langle \sigma v\rangle)\}$ are the values of the $J$-factors that maximize the likelihood for a given $m_\chi$ and $\langle \sigma v\rangle$.  A 95$\%$ C.L. upper limit is then defined by demanding $-\Delta\,\text {ln}\,{\cal L}(m_\chi, \langle \sigma v\rangle) \leq 2.71/2$.

\begin{figure}[t]
\centering
\includegraphics[width=0.55 \textwidth ]{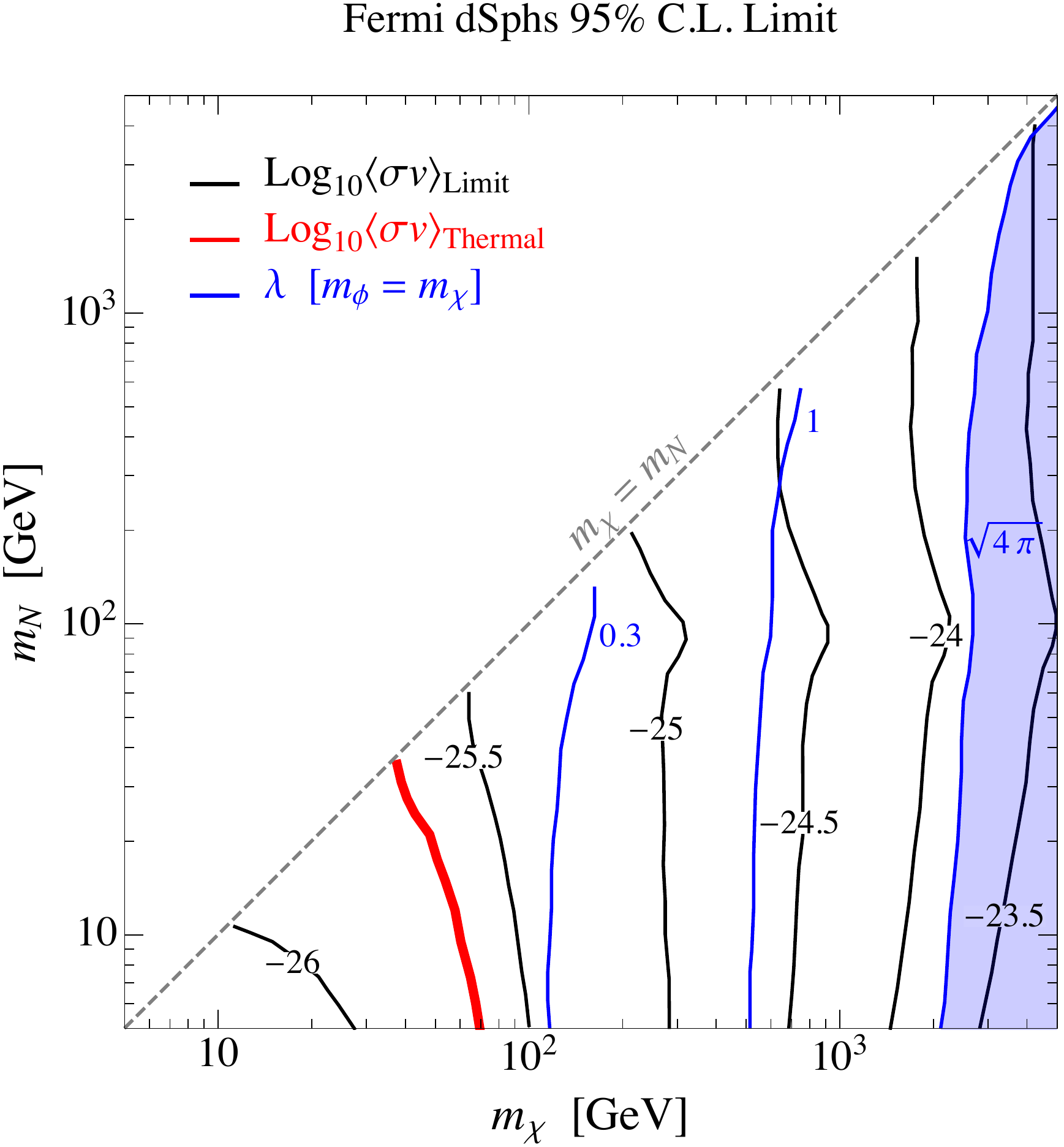}
\caption{\footnotesize 
Contours of the $95\%$ C.L. upper limit on $\log_{10}\left[\langle \sigma v \rangle/ ({\rm cm}^3\, {\rm s^{-1}})\right]$  in the $m_\chi - m_N$ plane (black curves) from {\it Fermi} observations of gamma-rays from the Milky Way dSphs. 
The thick (red) line indicates the region where the cross section limit is equal to the thermal relic value of Eq.~(\ref{eq:sigmav}). The vertical (blue) lines show the limits on the minimum value of the coupling constant $\lambda$. The shaded (blue) region indicates the perturbative unitarity bound. 
}
\label{fig:Dwarfcontour}
\end{figure}

We follow a similar approach to the {\it Fermi} prescription defined above, with one minor modification to speed up the numerical optimization. Rather than optimize over each of the 15 nuisance $J$-factors for each dSph, we introduce a single parameter, $\delta$, which represents the deviation of the $J$-factor of the dwarfs from their central values according to $\log_{10}(J_k) = \log_{10}(\bar J_k) + \delta\, \sigma_k$. 
Since no gamma-ray excess is observed in any indiviudual dSph, it is reasonable to expect that the fit tends to move all $J$-factors up or down simultaneously depending on the assumed values of $m_\chi$ and  $\langle\sigma v \rangle$, and this effect that is captured well by our $\delta$ prescription. 
As a validation, we have checked that our prescription reproduces the {\it Fermi} limits on DM annihilation in the $b \bar b$ channel~\cite{Ackermann:2015zua} at the 10-20$\%$ level throughout the entire mass range. 

Using the gamma ray spectra produced with the Monte-Carlo simulation described at the beginning of this section (examples are shown in Figure~\ref{fig:spectra}), we derive limits on the neutrino portal DM model for the channel $\chi \chi \rightarrow NN$. In Figure~\ref{fig:Dwarfcontour} we show contours of the 95$\%$ C.L. upper limit  on the annihilation cross section in the $m_\chi - m_N$ plane. 
The {\it Fermi} data from the Milky Way dSphs are able to probe thermal relic cross sections (\ref{eq:sigmav}) for $m_\chi \sim 40 - 80$ GeV as shown by the thick (red) line, depending on the mass of the RHN
\footnote{Our annihilation cross section limits are weaker than those dervied in Ref.~\cite{Campos:2017odj} by roughly a factor of two. We have not been able to find the source of the discrepancy, although it is perhaps possible to attribute the difference to the uncertainties in the dSph $J$-factors. We are grateful to Farinaldo Queiroz for correspondence on this issue.}.
 The vertical (blue) lines and the associated numbers show the limits on the minimum value of the coupling constant $\lambda$. The shaded (blue) region indicates the perturbative unitarity bound. 
In the region $m_W \lesssim m_N \lesssim m_Z$ the two body decay $N \rightarrow W \ell$ opens up and saturates the branching ratio, which is clearly seen in Figure~\ref{fig:Dwarfcontour}.

\subsection{Antiprotons}

Antiprotons ($\bar p$) have long been recognized as a promising indirect signature of DM. While DM annihilation typically produces equal numbers of protons and antiprotons, the astrophysical background flux of antiprotons is very small in comparison to that of protons. On the other hand, describing the production and propagation of these charged hadrons is a challenging task, and any statement regarding DM annihilation rests on our ability to understand the  associated astrophysical uncertainties. The Alpha Magnetic Spectrometer (AMS-02) experiment has provided the most precise measurements of the cosmic ray proton and antiproton flux to date~\cite{Aguilar:2016kjl}, and here we will explore the implications of this data on our neutrino portal DM scenario. Since DM annihilates to RHNs, which subsequently decay via $W$, $Z$, and Higgs bosons, the resulting cascade decay, showering and hadronization produce a variety of hadronic final states including antiprotons. AMS-02 will therefore provide an important probe of the model.

The propagation of antiprotons through the galaxy to earth is described by a diffusion equation for the distribution of antiprotons in energy and space (see, {\it e.g.}, Ref.~\cite{Boudaud:2014qra} and references therein). The transport is modeled in a diffusive region taken to be a cylindrical disk around the galactic plane and is affected by several physical processes. These include diffusion of the antiprotons through the turbulent magnetic fields, convective winds that impel antiprotons outward, energy loss processes, solar modulation, and a source term describing the production and loss of antiprotons. The source term accounts for astrophysical sources such as secondary and tertiary antiprotons, and antiproton annihilation with the interstellar gas, as well as primary antiprotons produced through DM annihilation. The propagation depends on a number of input parameters, and a set of canonical models, called MIN, MED, MAX are often employed~\cite{Donato:2005my}. 
The diffusion equation is solved assuming the steady state condition to find the flux of antiprotons from DM annihilation at earth, 
\begin{eqnarray}
\Phi_{{\bar p},\chi}(K)=\frac{v_{\bar p}}{4\pi}\left(\frac{\rho_{0}}{m_{\chi}}\right)^2 R(K)
\frac{1}{2} \langle\sigma v\rangle \frac{dN_{\bar p}}{dK},
\label{eq:antiflux}
\end{eqnarray}
where $dN_{\bar p}/dK$ is the kinetic energy ($K$) spectrum of antiprotons per DM annihilation, $v_{\bar p}$ is the antiproton velocity, and $\rho_{0}$ is the local DM density. The propagation function $R(K)$ accounts for the astrophysics of production and propagation, and we use the parameterization provided in Ref.~\cite{Cirelli:2010xx}.

\begin{figure}[t]
\centering
\includegraphics[width=0.55\textwidth ]{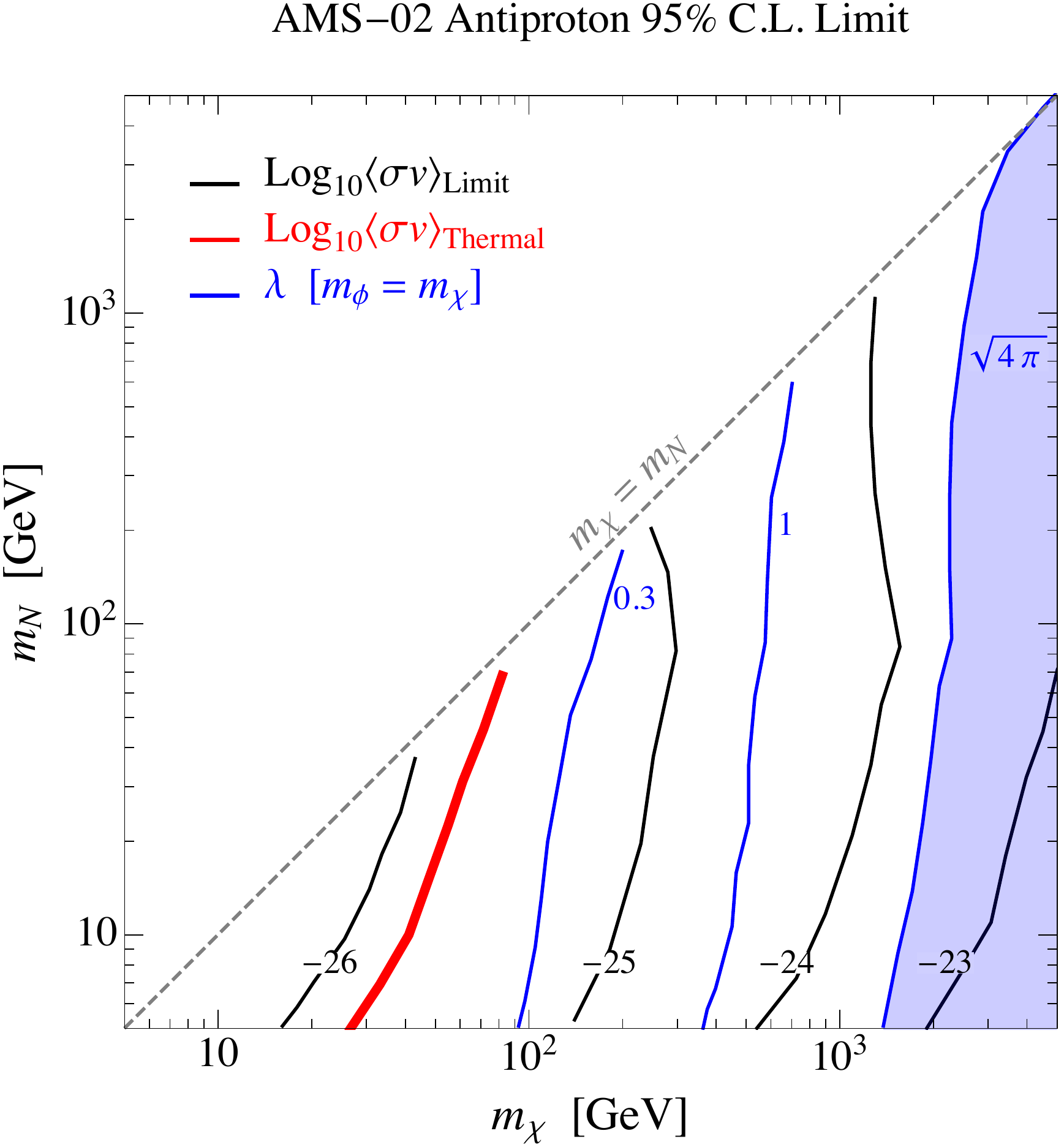}~
\caption{\footnotesize 
Contours of the  
upper limit on $\log_{10}\left[\langle \sigma v \rangle/ ({\rm cm}^3\, {\rm s^{-1}})\right]$  in the $m_\chi - m_N$ plane (black curves) from the AMS-02 measurement of the antiproton-to-proton flux ratio~\cite{Aguilar:2016kjl}. 
The thick (red) line indicates the region where the cross section limit is equal to the thermal relic value of Eq.~(\ref{eq:sigmav}). The vertical (blue) lines show the limits on the minimum value of the coupling constant $\lambda$. The shaded (blue) region indicates the perturbative unitarity bound. 
}
\label{fig:PROcontour}
\end{figure}

AMS-02 has provided precise measurements of the proton flux, $\Phi_p (K)$~\cite{Aguilar:2015ooa},
 and the antiproton-to-proton flux ratio, $r(K)$~\cite{Aguilar:2016kjl}, which can be used to place constraints on DM annihilation. 
To proceed, we require an estimate of the secondary background antiproton flux originating from astrophysical sources. 
For this purpose we use the best-fit secondary flux, $\Phi_{\bar p,\text{bkg}}(K)$, from \cite{Giesen:2015ufa}, which provides an acceptable fit to the AMS-02 data. 
With the total antiproton flux, $\Phi_{\bar p,\text{tot}}(K,m_{\chi},\langle\sigma v\rangle)= \Phi_{\bar p,\text{bkg}}(K)+\Phi_{\bar p,\chi}(K,m_{\chi},\langle\sigma v\rangle)$, and the measured proton flux from AMS-02, $\Phi_p(K)$, in hand, we form the ratio of these two fluxes and fit it to the observed ratio. The test statistic is 
\begin{eqnarray}
\chi^2(m_{\chi},\langle\sigma v\rangle)=\sum_i\frac{[r(K_i)-(\Phi_{\bar p,\text{tot}}(K_i,m_{\chi},\langle\sigma v\rangle)/\Phi_p(K_i))]^2}{\sigma_i^2},
\label{eq:chi}
\end{eqnarray}
where $i$ runs over energy bins, and $\sigma_i$ is the reported uncertainty of the flux ratio~\cite{Aguilar:2016kjl}. 
Following Ref.~\cite{Giesen:2015ufa}, we define a limit on $\langle\sigma v\rangle$ as a function of $m_\chi$, $m_N$ according to the condition
\begin{eqnarray}
\chi^2(m_{\chi},\langle\sigma v\rangle)-\chi_0^2\leq 4.
\label{eq:bound}
\end{eqnarray}
where $\chi_0^2$ is the best fit chi-squared statistic assuming no primary DM antiproton source from Ref.~\cite{Giesen:2015ufa}. The limit is derived under the assumption of a Einasto profile and using the MED propagation scheme. Contours of the limit on the annihilation cross section in the $m_\chi-m_N$ plane are displayed in Figure~\ref{fig:PROcontour}. For DM masses in the range of 20 - 80 GeV, AMS-02 is able to probe the thermal cross section Eq.~(\ref{eq:sigmav}), as indicated by the thick (red) line.
The vertical (blue) lines show the limits on the minimum value of the coupling constant $\lambda$. The shaded (blue) region indicates the perturbative unitarity bound. 
It is important to note again that there are significant uncertainties associated with the DM halo profile and the propagation scheme, which can lead to a variation in the cross section limits by one order of magnitude or more~\cite{Giesen:2015ufa}. Note that for a fixed $m_\chi$, the limits in Figure~\ref{fig:PROcontour} become stronger as $m_N$ is increased. This is because for fixed $m_\chi$, heavier RHNs tend to produce more low energy antiprotons (see Figure~\ref{fig:spectra}). However, the ratio $r(K)$ shows good agreement with the astrophysical background model at low value of kinetic energy $K$ and a slight excess at larger values of $K$, explaining the behavior seen in Figure~\ref{fig:PROcontour}.

\begin{figure}[t]
\centering
\includegraphics[width=0.55 \textwidth ]{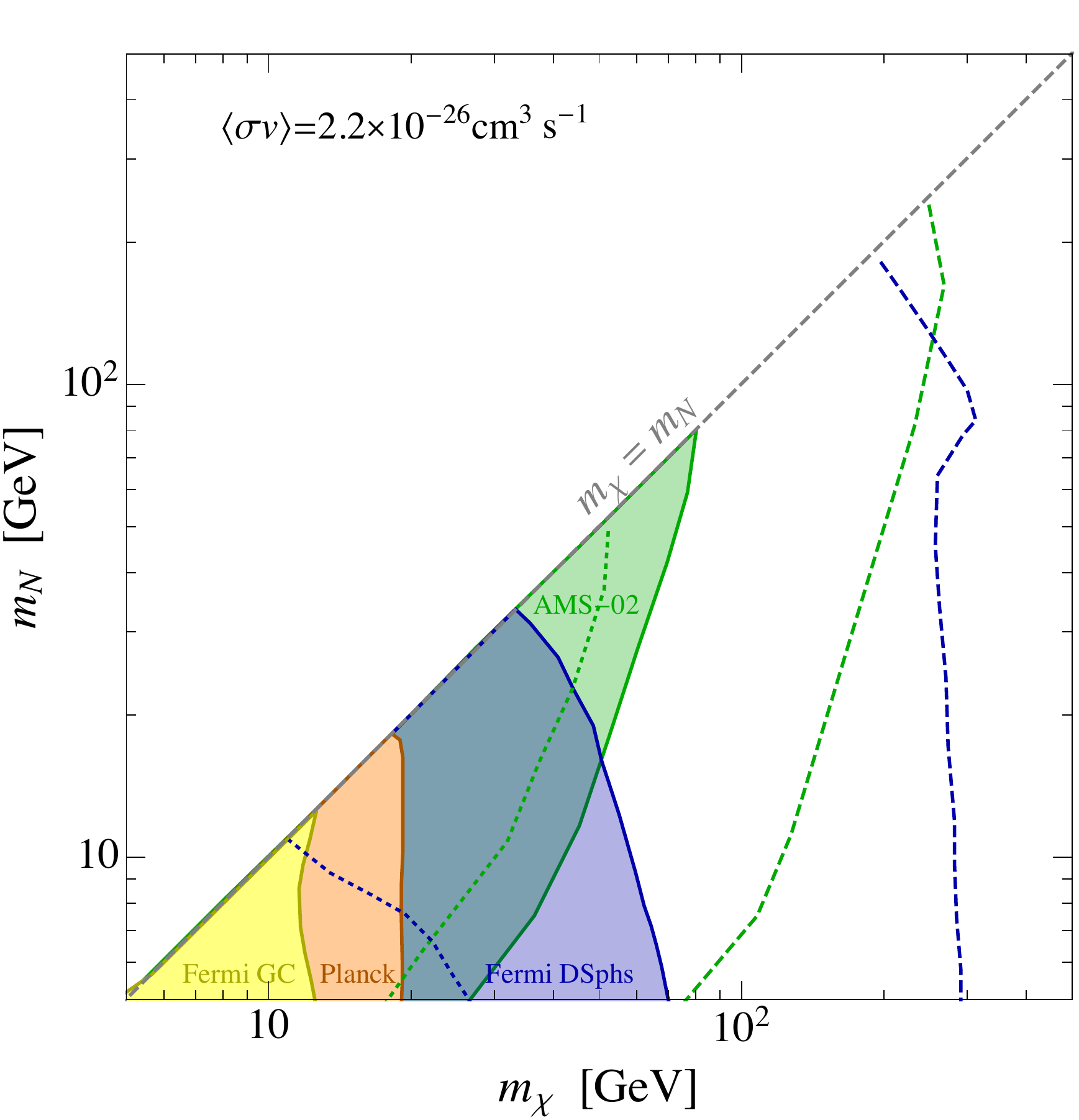}
\caption{\footnotesize 
Constraints on the 
neutrino portal DM model in the
in the $m_\chi - m_N$ plane from the CMB ({\it Planck}), Galactic Center gamma rays ({\it Fermi}), dSphs gamma rays ({\it Fermi}), and antiprotons (AMS-02). 
A thermal annihilation cross section $\langle \sigma v \rangle = 2.2 \times 10^{-26}$ cm$^3$\ s$^{-1}$ is assumed throughout. 
See the text and Figures~\ref{fig:CMBcontour},\ref{fig:GCRcontour},\ref{fig:Dwarfcontour},\ref{fig:PROcontour} for further details. 
Dotted and dashed lines illustrate the impact of DM-related astrophysical uncertainties. 
}
\label{fig:combined}
\end{figure}

\subsection{Summary of limits and future prospects}
\label{sec:summary}

In Figure~\ref{fig:combined} we show the combined limits on the neutrino portal DM model for the case in which the annihilation cross section is fixed to the thermal value, $\langle \sigma v \rangle = 2.2 \times 10^{-26}$ cm$^3$ s$^{-1}$. Constraints from {\it Planck} CMB measurements, {\it Fermi} observations of gamma-rays from the Galactic Center and dSphs, and AMS-02 antiproton measurements are shown. We remind the reader that the {\it Fermi} Galactic Center limits are derived for the choice of an NFW halo profile, while the AMS-02 antiproton limits are based on an Einasto profile and MED propagation scheme. 
Under the stated assumptions, we conclude that thermal annihilation is constrained for DM masses up to $50-70$ GeV depending on RHN mass. 
AMS-02 provides the best probe in the case $m_N \lesssim m_\chi$, while {\it Fermi} dSphs provides the superior constraint for $m_N \ll m_\chi$. We have also illustrated the impact of astrophysical uncertainties on the antiproton and dSphs limits in Figure~\ref{fig:combined}. For antiproton constraints, we show Burkert profile and MED propagation (green dotted line) and  Einasto profile and MAX propagation (green dashed line). For dSphs, we show $\log_{10}(J_k) = \log_{10}(\bar J_k) - 2\sigma_k$ (blue dotted line) and $\log_{10}(J_k) = \log_{10}(\bar J_k) +2\sigma_k$ (blue dashed line).

There are several other notable indirect DM searches that we wish to comment on here. AMS-02 has provided detailed measurements of the cosmic ray positron spectrum~\cite{Aguilar:2014mma}. Much attention has been paid to these results (and those of its forerunner PAMELA~\cite{Adriani:2013uda}) due to the observation of a striking rise in the fractional positron flux, which potentially points to a new primary source of positrons. While it is true that DM annihilation in our scenario produces a significant positron flux, the cross section limits from {\it Fermi} dSphs gamma rays and AMS-02 antiproton observations are expected to be stronger than those from AMS-02 positron measurements by an order of magnitude or more, and thus we have chosen to focus on these stronger tests. 

Another well-known indirect DM probe is high energy neutrinos from DM annihilation in the sun, which can be probed with the IceCube experiment~\cite{Aartsen:2016zhm}. But under the minimal assumption of typical seesaw values for the neutrino Yukawa coupling~(see Eq.~(\ref{eq:yukawa})) the DM-nucleon scattering rate will be too small to allow for the efficient capture of DM in the sun, so we do not consider this possibility further.

Along with the continuum gamma-ray signatures studied here, there is also the possibility of a harder gamma-ray spectral feature that arises from the radiative decay $N \rightarrow \gamma \nu$~\cite{Ibarra:2016fco}.
This signature will be relevant in the region $m_\chi \sim m_N$, $m_N \lesssim 50$ GeV. For the benchmark thermal relic cross section, there are already relevant limits in this region from AMS-02 (see Figure~\ref{fig:combined}), which however are subject to sizable astrophysical uncertainties. In that regard, the spectral ``triangle'' feature would provide a complementary probe. On the one hand, the hard spectral feature has the advantage of being more easily discernible over the power law background, while at the same time it is expected that the overall rate will be significantly less than the gamma-ray continuum signal due to its radiative origin. A full quantitative study of this signature goes beyond our scope here and we refer the reader to Ref.~\cite{Ibarra:2016fco} for further details.

As we have demonstrated, the data collected so far by {\it Fermi}-LAT already leads to stringent limits on DM parameter space, and the sensitivity will improve significantly in the coming years. The projected sensitivities for 10 and 15 years of data taking has been studied in detail by the collaboration in Ref.~\cite{Charles:2016pgz}. The fast discovery of new dSphs is the primary upcoming change in dSph targeted DM searches. The identification of new dSph candidates by the Dark Energy Survey (DES)~\cite{Abbott:2005bi} over the past two years, if confirmed, will double the number of known dSphs. Following on important discoveries of the Sloan Digital Sky Survey (SDSS)~\cite{York:2000gk}, which covered $1/3$ of the sky and discovered 15 ultra-faint dSphs, surveys like DES and especially the Large Synoptic Survey Telescope (LSST)~\cite{Ivezic:2008fe} will cover complementary regions of the sky which are expected to discover potentially ${\cal O}(100)$ dSphs. Ref.~\cite{Charles:2016pgz} takes 60 total dSphs as an estimate of the number of dSphs that can be used for LAT searches. They find that the sensitivity of searches targeting dwarf galaxies will improve faster than the square root of observing time. 
Following Ref.~\cite{Charles:2016pgz} we expect an improvement on the cross section limit from {\it Fermi}-LAT 15 years dSph observations by a factor of a few, which will probe thermal relic DM with masses $m_\chi \gtrsim 100$ GeV in the neutrino portal DM scenario.

Due to their large effective areas, ground-based imaging air Cherenkov telescopes (IACTs), such as 
H.E.S.S.~\cite{Abdallah:2016ygi}, 
VERITAS~\cite{Smith:2013tta}, and MAGIC~\cite{Aleksic:2013xea}, and in the future CTA~\cite{Doro:2012xx}
 and 
HAWC~\cite{Abeysekara:2014ffg}, are well suited to search for higher energy gamma rays originating from heavy DM annihilation. In particular, H.E.S.S. has presented a search for DM annihilation towards the Galactic Center using 10 years of data~\cite{Abdallah:2016ygi}. 
Assuming a cuspy NFW or Einasto profile the search sets the strongest limits on TeV mass DM that annihilates to $WW$ or quarks, and almost reach thermal annihilation rates. Taken at face value, the H.E.S.S. limits are indeed stronger than the {\it Fermi} dSphs limits for DM masses above a few hundred GeV, but are however less robust due to the inherent astrophysical uncertainties associated with the central region of the Milky way, both in terms of conventional gamma-ray sources and the DM distribution. The H.E.S.S data is not publicly available, so unfortunately we are not able to properly recast their limit. 
However, for a fixed DM mass, the continuum photon spectrum produced in our model from $\chi \chi \rightarrow NN$ is qualitatively similar to the spectrum produced by $\chi\chi \rightarrow WW$. We can therefore obtain a rough estimate of the H.E.S.S. sensitivity by translating their limits in the $WW$ channel to our parameter space
The H.E.S.S. limits are approaching the canonical thermal relic annihilation rate for DM masses around 1 TeV.

In the future, the Cherenkov Telescope Array (CTA) will be able to further probe heavy TeV-scale DM annihilation, with the potential to improve by roughly an order of magnitude in cross section sensitivity over current instruments depending on the annihilation mode and DM mass. Here we estimate the sensitivity of future CTA gamma-ray observations of the Galactic Center using a ``Ring'' method technique~\cite{Moulin:2013lma}. Our projections are based on a simplified version of the analysis carried out in Ref.~\cite{Silverwood:2014yza} that we now briefly describe. 
The analysis begins with the definition of signal (referred to as ``ON'') and background (``OFF'') regions. 
A binned Poisson likelihood function is constructed in order to compare the DM model $\pmb{\mu}$ to a (mock) data set $\textbf{\textit{n}}$:
\begin{eqnarray}
{\cal L}(\pmb{\mu}\rvert \textbf{\textit{n}})=\prod_{i,j}\frac{\mu_{ij}^{n_{ij}}}{n_{ij}!}e^{-\mu_{ij}}.
\label{eq:CTAlike}
\end{eqnarray}
where $\mu_{ij}$ is the predicted number of events for a given model $\pmb{\mu}$ in the $i$th energy bin and the $j$th region of interest, corresponding to ON ($j=1$) and OFF ($j = 2$) regions.  These model predictions are compared to the corresponding observed counts $n_{ij}$. We use 15 logarithmically-spaced energy bins, extending from 25 GeV to 10 TeV. 
The number of gamma-ray events predicted by each model consists of three components: a DM annihilation signal, an isotropic cosmic-ray (CR) background, and the Galactic diffuse emission (GDE) background:
\begin{eqnarray}
\mu_{ij}=\mu_{ij}^{\text{DM}}+\mu_{ij}^{\text{CR}}+\mu_{ij}^{\text{GDE}}.
\end{eqnarray}
The details for the regions of interest that 
have been used in our analysis, including the corresponding solid angles and $J$-factors, can be found in Ref.~\cite{Doro:2012xx}. 
Furthermore, we have used the effective area produced by MPIK group \cite{Bernlohr:2012we} and fixed the time of observation to be 100 hours.

\begin{figure}[t]
\centering
\includegraphics[width=0.55 \textwidth ]{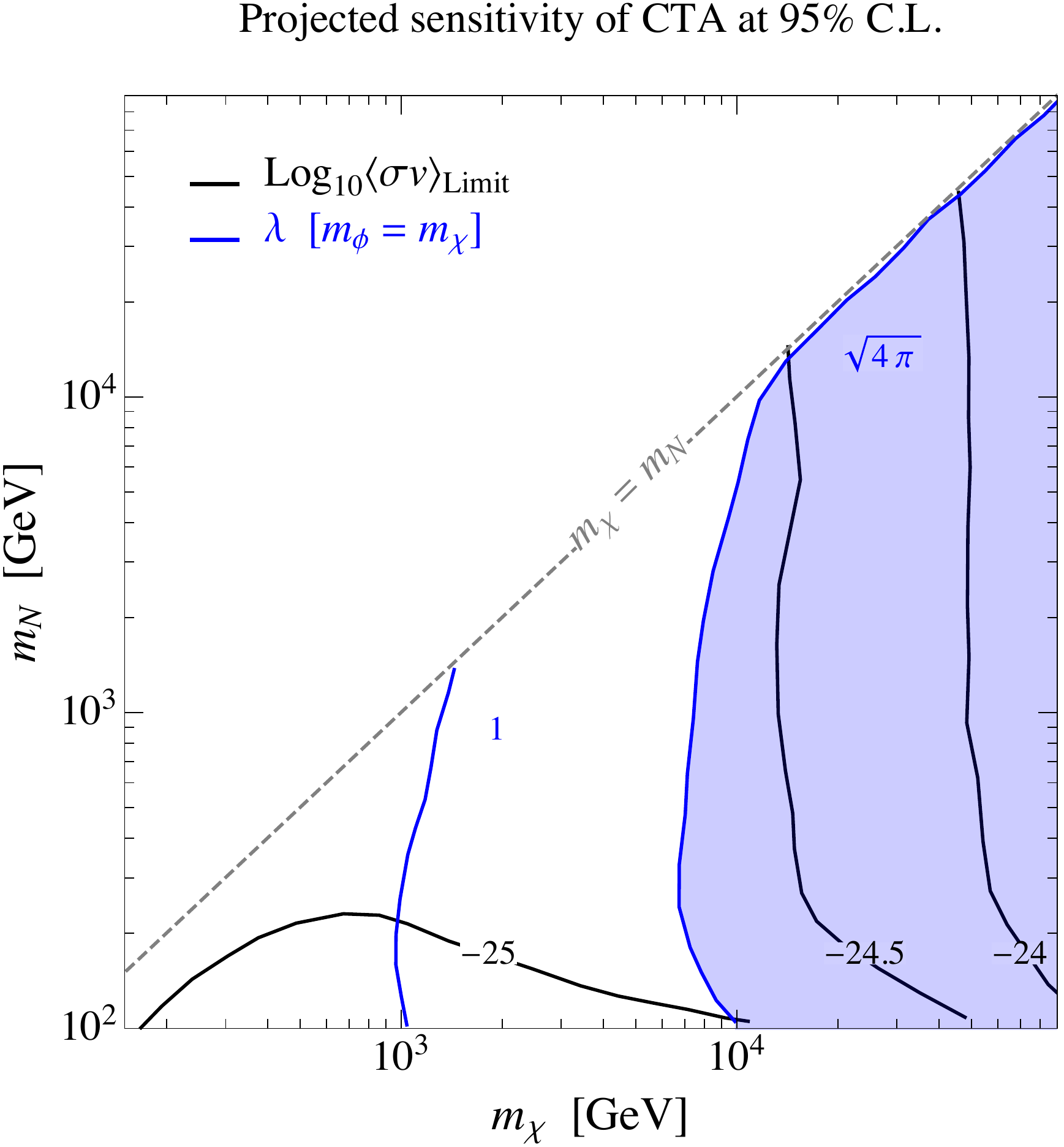}
\caption{\footnotesize 
Contours of the $95\%$ C.L. projected sensetivity on $\log_{10}\left[\langle \sigma v \rangle/ ({\rm cm}^3\, {\rm s^{-1}})\right]$ in the $m_\chi - m_N$ plane (black curves) from CTA $\gamma$-ray observations of the Galactic Center using Ring method, assuming 100hr of observation \cite{Silverwood:2014yza}.
}
\label{fig:CTA}
\end{figure}

We account for differential acceptance uncertainties (i.e. acceptance variations across different energy bins and regions-of-interest) by rescaling the predicted signals $\mu_{ij}$ by parameters $\alpha_{ij}$ and profiling the likelihood over their values. Following Ref.~\cite{Silverwood:2014yza} we assume Gaussian nuisance likelihoods for all $\alpha$ with respective variance $\sigma_{\alpha}^2$ independent of $i$ and $j$. Our limits correspond to differential acceptance uncertainties of 1\%. 
The mock data $\textbf{\textit{n}}$ we employ includes a fixed isotropic cosmic-ray background component in all bins, and no signal from DM annihilation. 
We derive 95$\%$ CL upper limits (sensitivity) on the annihilation cross-section $\langle \sigma v \rangle$ in the usual way by requiring $-\Delta \ln {\cal L}  \leq 2.71/2$. Our projections are shown in Figure~\ref{fig:CTA}. We have not included systematic uncertainties for the background components, which can be as large as order one and thus significantly degrade the CTA sensitivity. However, this can be partially overcome through a more sophisticated morphological analysis, which leverages the shape differences between the galactic diffuse emission and DM signal~\cite{Silverwood:2014yza}. In the end, we expect that Figure~\ref{fig:CTA} provides a reasonable ballpark estimate of the CTA sensitivity, which can improve over H.E.S.S. by a factor of a few to ten in the 100 GeV - TeV DM mass range. We expect {\it Fermi} dSphs observations to provide superior limits for lower mass DM, $m_\chi \lesssim 100$ GeV.

\section{Galactic Center gamma ray excess  interpretation }
\label{sec:excess}

As  mentioned in Section \ref{sec:GC}, various analyses of {\it Fermi}-LAT data show a spherically symmetric excess of gamma rays coming from the central region of the Milky Way peaking in the 1-3 GeV energy range~\cite{Goodenough:2009gk,Hooper:2010mq,Daylan:2014rsa,Calore:2014xka,TheFermi-LAT:2015kwa}. Since DM annihilation to RHNs abundantly produces gamma rays, it is interesting to explore a possible interpretation of this excess in the context of the neutrino portal DM model. In fact, this possibility was previously investigated in Ref.~\cite{Tang:2015coo}, which found that DM annihilation to RHNs could indeed provide a good fit to the Galactic Center excess. Here we will additionally confront this interpretation with existing constraints from other indirect probes, and notably {\it Fermi} gamma-ray observations from dSphs and AMS-02 antiproton observations. 

\begin{figure}[t]
\centering
\includegraphics[width=0.46 \textwidth ]{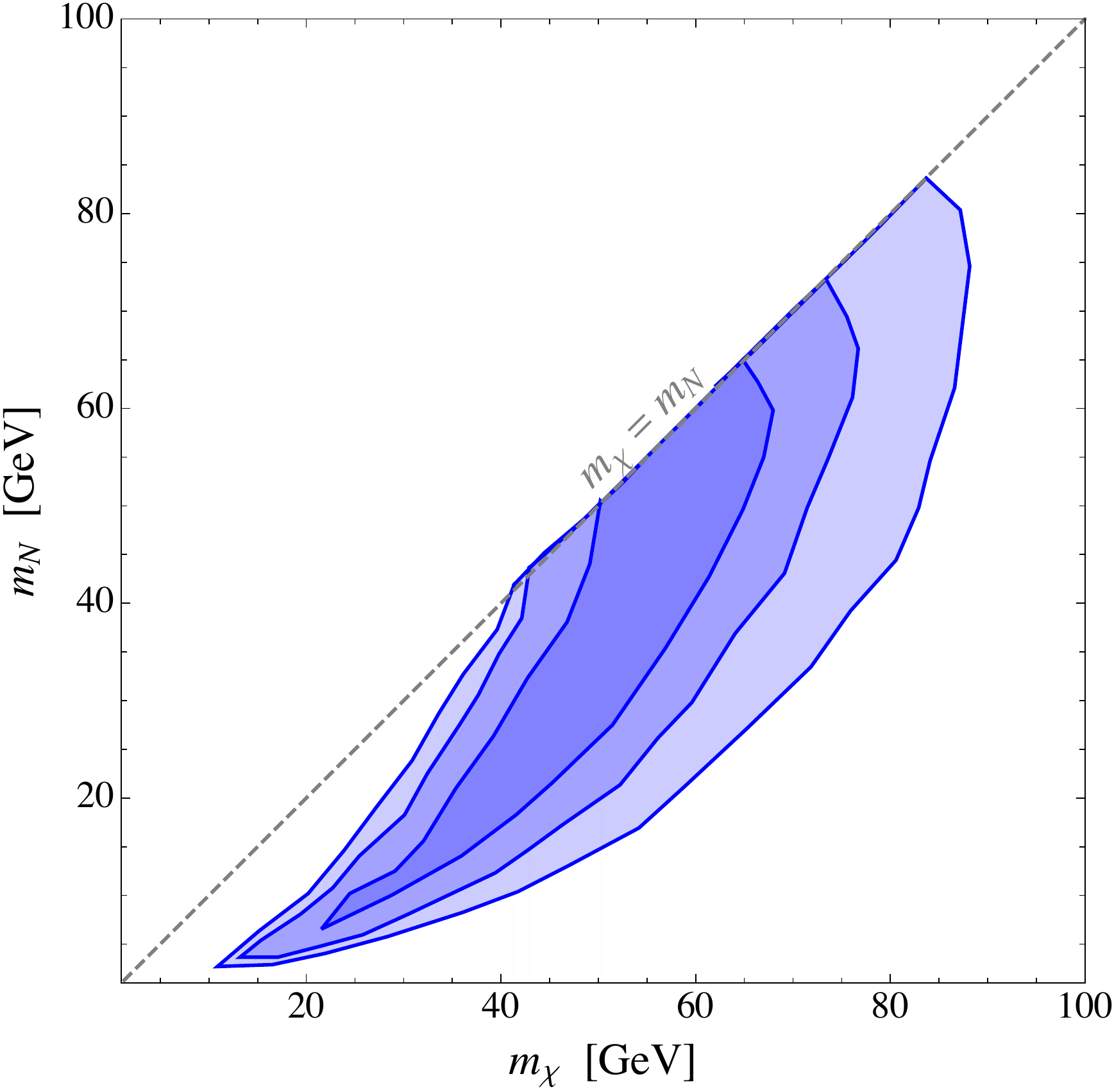}
\includegraphics[width=0.45 \textwidth ]{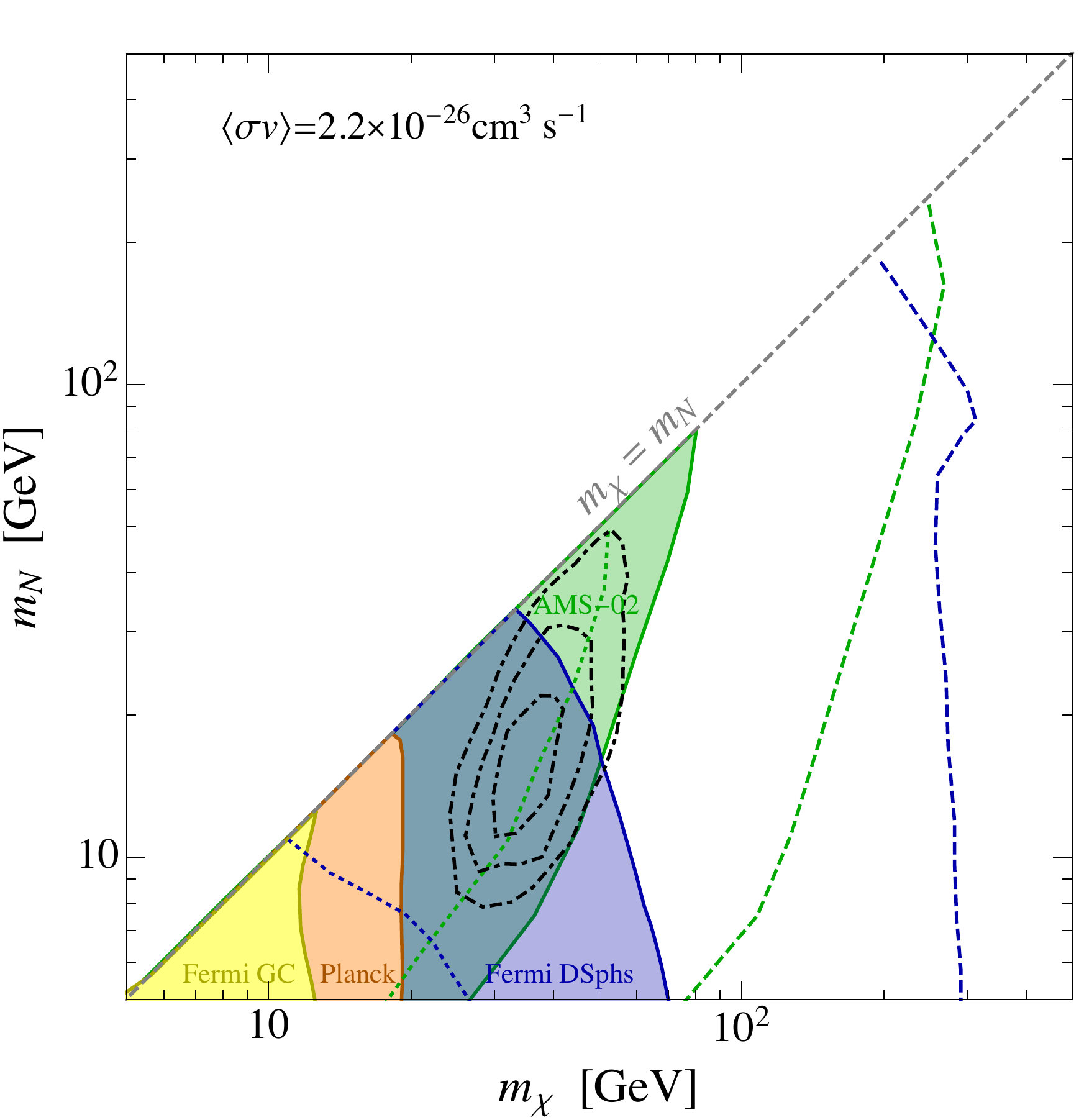}
\caption{\footnotesize Interpretation of the Galactic Center gamma ray excess. The left panel displays the $1\sigma,2\sigma$, and $3\sigma$ preferred regions in the $m_\chi - m_N$ plane, with the best-fit point of $\{\langle\sigma v\rangle=3.08\times 10^{-26}\ \text{cm}^3\ \text{s}^{-1},m_{\chi}=41.3\ \text{GeV}, m_N=22.6 \ \text{GeV}\}$ with $\chi^2=14.12$ for 23 degrees of freedom.  The right panel shows the best-fit region for the case of a fixed thermal annihilation cross section, $\langle \sigma v \rangle = 2.2 \times 10^{-26}$ cm$^3$ s$^{-1}$, as well as the existing limits from {\it Planck} CMB, {\it Fermi} dSphs, and AMS-02 antiproton observations.
}
\label{fig:excess}
\end{figure}

We fit the neutrino portal DM model parameters to the Galactic Center excess spectrum given in Ref.~\cite{Calore:2014xka}. We adopt Navarro-Frenk-White (NFW) profile with $\gamma=1.2$. Following \cite{Calore:2014xka} we define the $\chi^2$ as
\begin{eqnarray}
\chi^2(\mbox{\boldmath$\theta$})=\sum_{ij}\left[\Phi_i(\mbox{\boldmath$\theta$})-\left(\Phi_i\right)_{\text{obs}}\right]\cdot\Sigma_{ij}^{-1}\cdot\left[\Phi_j(\mbox{\boldmath$\theta$})-\left(\Phi_j\right)_{\text{obs}}\right],
\label{eq:chiexcess}
\end{eqnarray}
where \mbox{\boldmath$\theta$}=$\{\langle\sigma v\rangle, m_{\chi},m_N\}$, $\Phi_i\ ( (\Phi_i)_{\text{obs}})$ is the predicted (observed) $\gamma$-ray flux~(see Eq.~(\ref{eq:photonflux})) in the $i^{th}$ energy bin, and $\Sigma$ is the covariance matrix. We find that the best-fit point is 
$\{\langle\sigma v\rangle=3.08\times 10^{-26}\ \text{cm}^3\ \text{s}^{-1},m_{\chi}=41.3\ \text{GeV}, m_N=22.6 \ \text{GeV}\}$ with $\chi^2=14.12$ for 23 degrees-of-freedom. 
Figure~\ref{fig:excess} displays $1\sigma,2\sigma$, and $3\sigma$ CL regions in the $m_N - m_\chi$ parameter space. 
We see that neutrino portal DM can provide an acceptable fit over a significant range of mass parameters. 

Next, we would like to confront this interpretation with the other constraints derived in Section~\ref{sec:indirect}. To this end, we perform the Galactic Center excess while fit fixing the annihilation cross section to its  thermal value, and overlay the limits derived from {\it Planck} CMB, {\it Fermi} dSphs, and AMS-02 antiproton observations. The result is displayed in the right panel of Figure~\ref{fig:excess}. We see that this interpretation faces some tension with limits from dSphs and antiprotons. However, it is too early to conclude from this analysis that the DM interpretation of the excess is not viable given the significant astrophysical uncertainties in the local DM density, dSphs DM densities, and the modeling of the antiproton propagation.

\section{Beyond the minimal scenario}
\label{sec:beyond}

We have explored what is perhaps the simplest scenario of neutrino portal DM. The primary probe of this model comes from indirect detection, and we have presented a comprehensive picture of the current constraints. However, it is possible that the neutrino mass model is more complex than the simplest Type-I seesaw, or that there are additional interactions of the scalar mediator with the Higgs, in which case a much richer phenomenology is possible. In this section we will highlight some of these possibilities. 

\subsection{Large neutrino Yukawa coupling}

Taking the naive seesaw relation in Eq.~(\ref{eq:seesaw}) as a guide, one generally expects very small active-sterile mixing angles, $\theta \sim \sqrt{m_\nu/m_N}\simeq 10^{-6}\times \left(m_N/{100 \, \rm GeV}\right)^{-1/2}$, suggesting poor prospects for direct detection and accelerator experiments. However, the neutrino Yukawa coupling and active sterile-mixing angle can be  much larger if one goes beyond the simplest Type-I seesaw. For example, in the inverse seesaw model~\cite{Mohapatra:1986aw}, the RHNs are pseudo-Dirac fermions, with splitting governed given by a small Majorana mass. The SM neutrino masses are light due to the same small Majorana mass, while the Yukawa coupling can in principle be as large as $y \sim 0.1$, while being compatible with experimental constraints. 

Such large Yukawa couplings not only offer increased chances to probe the RHNs directly (see, {\it e.g.}, Ref.~\cite{Atre:2009rg,Deppisch:2015qwa} for a revew), but will also enhance the detection prospects of the DM sector. For instance, one can induce sizable DM couplings to the $Z$ and Higgs boson at one loop that mediate large scattering rates with nuclei, which is relevant for direct detection experiments and capture of DM in the sun. One can also potentially produce the RHNs directly in accelerator experiments. 

This also opens up the possibility for the RHN to be heavier than the dark sector particles, while still having a thermal cosmology. Due to the large mixing angle, it is possible for DM to annihilate efficiently into light active neutrinos, and furthermore the DM may annihilate to other SM particles through the loop-induced $Z$ and $h$ couplings. We refer the reader to Refs.~\cite{Bertoni:2014mva,Macias:2015cna,Gonzalez-Macias:2016vxy} for recent investigations of these issues. 

\subsection{Higgs portal coupling}

The scalar particle $\phi$ can couple to the Higgs portal at the renormalizable level 
\begin{equation}
{\cal L}\supset \frac{\lambda_{\phi H}}{2} \phi^2 |H|^2.
\label{eq:higgsportal}
\end{equation}
We have so far assumed that this coupling is small. The reason we have made this assumption is primarily for simplicity, as  then the phenomenology and cosmology is solely dictated by the neutrino portal link to the SM. However, this assumption can certainly be questioned. 

Restricting to the fields and interactions of our scenario in Eq.~(\ref{eq:type1L}), we observe that the Higgs portal coupling (\ref{eq:higgsportal}) will be induced at one loop with strength of order $\lambda_{\phi H} \sim \lambda^2 y^2/16 \pi^2$, which is very small due to the small neutrino Yukawa coupling. Still, one may expect unknown UV physics to generically induce a larger coupling. This is because there is no enhanced symmetry in the limit $\lambda_{\phi H} \rightarrow 0$, and so even though the operator (\ref{eq:higgsportal}) is marginal, we cannot rely on technical naturalness ensure a small value without further information about the UV physics. That being said, one can certainly imagine completions in which the Higgs portal coupling is suppressed. For example if $\phi$ is a composite scalar state of some new strong dynamics, then the Higgs portal operator would fundamentally be a higher dimension operator and could be therefore be naturally suppressed. 

Another good reason to consider the Higgs portal operator is that it provides additional opportunities to probe the dark sector in experiment. A one loop coupling of the DM to the Higgs will be induced and this can mediate scattering of DM with nuclei, or invisible decays of the Higgs to DM~\cite{Macias:2015cna,Gonzalez-Macias:2016vxy,Escudero:2016tzx,Escudero:2016ksa}.

An even more distinctive signature at colliders can arise if the Higgs could decay into a pair of light scalars, $h\rightarrow \phi\phi$. These scalars, once produced would then cascade decay via $\phi \rightarrow N \chi$. The resulting RHN $N$, being lighter than the $W$ boson, will have a macroscopic decay length and could leave a striking displaced vertex signal (see, {\it e.g.},~\cite{Izaguirre:2015pga}). The signature would thus be an exotic Higgs decay with two displaced vertices.

\section{Summary and Outlook}
\label{sec:outlook}

In this paper, we have investigated a simple model of neutrino portal DM, in which the RHNs simultaneously generate light neutrino masses via the Type-I seesaw mechanism and mediate interactions of DM with the SM. The model, presented in Section~\ref{sec:model}, is quite minimal and contains a dark sector composed of a fermion $\chi$ (the DM candidate) and scalar $\phi$, along with the RHN $N$. 
Given the generic expectation of tiny neutrino Yukawa couplings, testing this model with direct detection or accelerator experiments is likely to be challenging. However, it is possible in this model that DM efficiently annihilates to RHNs, which allows for a number of indirect probes of this scenario. 

We have carried out an extensive characterization of the indirect detection phenomenology of the neutrino portal DM scenario in Section \ref{sec:indirect}.
Restricting to an experimentally and theoretically viable mass range,  1 GeV $\lesssim m_N  < m_\chi \lesssim 10 $ TeV, we have derived the constraints on the $\chi \chi \rightarrow NN$ annihilation cross section from {\it Planck} CMB measurements, {\it Fermi} gamma-ray observations from the Galactic Center and from dSphs,  and AMS-02 antiproton observations. Currently, the dSphs and antiproton measurements constrain DM masses below 50 GeV for thermal annihilation rates. In the future, {\it Fermi} dSphs observations will be able probe DM masses above the 100 GeV range for thermal cross sections, while CTA will be able to approach thermal cross section values for DM masses in the 100 GeV - 1TeV range. 

This model can also provide a DM interpretation of the {\it Fermi} Galactic Center gamma ray excess as discussed in Section \ref{sec:excess}.  We have verified that the predicted spectrum of gamma rays is compatible with the observed excess for RHN and DM masses in the $20 - 60$ GeV range and annihilation rates close the the thermal value. However, we have also shown that this interpretation faces some tension with the existing constraints from {\it Fermi} dSphs and AMS-02 antiprotons (subject of course to various astrophysical uncertainties). It will be interesting to see how this situation develops as {\it Fermi} and AMS-02 collect more data. However, at least in the simplest model explored here, it will be challenging to find complementary probes outside of indirect detection. 

It is possible that the neutrino mass generation mechanism is more intricate than the simplest Type-I seesaw, as discussed in Section \ref{sec:beyond}. If so, the implications for neutrino portal DM could be dramatic, particularly if the neutrino Yukawa coupling is large, as this could lead to direct detection prospects, accelerator probes, and new annihilation channels. Additionally, it is possible in this scenario for additional Higgs portal couplings to be active, which could yield further phenomenological handles.

Portals provide a simple and predictive theoretical framework to characterize the allowed renormalizable interactions between the SM and DM. Furthermore, the existence of neutrino masses already provides a strong hint that the neutrino portal itself operates in nature. These two observations provide a solid motivation for testing the neutrino portal DM scenario, both through the generic indirect detection signals investigated in this paper, and also the additional signals present in more general models. It is worthwhile to broadly explore these scenarios and their associated phenomenology in detail, and we look forward to further progress in this direction in the future. 


 \subsubsection*{\bf Acknowledgements}
We thank David McKeen and Satyanarayan Mukhopadhyay for helpful discussions. We are also grateful to Farinaldo Queiroz for correspondence, and to Roberto Ruiz de Austri for pointing out an inconsistency in our antiproton spectrum in the first version of this paper. 
The work of BB and BSE is supported in part by the U.S. Department of Energy under grant No. DE-SC0015634, and in part by PITT PACC. 
The work of TH and BSE is supported in part by the Department of Energy under Grant No. DE-FG02-95ER40896, and in part by PITT PACC. 
We would also like to thank the Aspen Center for Physics for hospitality, where part of the work was completed. 
The Aspen Center for Physics is supported by the NSF under Grant No. PHYS-1066293.

\appendix
\numberwithin{equation}{section}

\section{Boosted spectrum for massive particles}
Consider first a particle of mass $m$ with a normalized monoenergetic and isotropic spectrum $f(E)$ in frame $O$ with energy $E_0$, i.e., 
\begin{eqnarray}
f_0(E)=\delta (E-E_0),\,\,\,\,\,\ \int_m ^\infty dE\ f_0(E)\, =1.
\end{eqnarray}
We wish to find the spectrum in a boosted frame $O'$.
In general there will be an angle $\theta$ between the  boost velocity $\boldsymbol{\beta}$ and the particle momentum, such that the 
the energy $E'$ in $O'$ is related to energy $E$ in $O$, as  
\begin{eqnarray}
E'=\gamma(E- \, \beta \, p \cos\theta ).
\label{eq:boosted-energy}
\end{eqnarray}
where $p = \sqrt{E^2-m^2}$.
Using Eq.~(\ref{eq:boosted-energy}) and averaging over the angle $\theta$ under the assumption of isotropy, one can show that the energies are uniformly distributed in $O'$ according to the ``box'' spectrum:
\begin{eqnarray}
f'_0(E') = \frac{1}{2\beta\gamma \, p_0}\,\theta[E'-\gamma(E_0-\beta \, p_0 )]\,\theta[\gamma(E_0+\beta \, p_0)-E'].
\label{eq:boost-mono}
\end{eqnarray}

We can use this result (\ref{eq:boost-mono}) to boost a general isotropic energy spectrum $f(E)$ observed in $O$, that in particular is not necessarily monoenergetic.  Starting from the normalization condition, we have
\begin{eqnarray}
1 & = & \int_m ^\infty dE\,f(E)
 =  \int_m ^\infty dE_0\,f(E_0)\left[\int_m^\infty dE\,\delta (E-E_0)\right],
\end{eqnarray}
where in the last step we have inserted the identity and changed the order of integration. 
The quantity in brackets is simply a monoenergtic spectrum with energy $E_0$ that was already considered above. 
Using Eq. (\ref{eq:boost-mono}), it is straightforward to derive the boosted spectrum in the frame $O'$:
\begin{eqnarray}
f'(E')=\int_{\gamma(E'-\beta\sqrt{E'^2-m^2})} ^{\gamma(E'+\beta\sqrt{E'^2-m^2})} \frac{dE}{2\beta\gamma \sqrt{E^2-m^2}}\,f(E).
\end{eqnarray}



\begin{thebibliography}{99}


\bibitem{Jungman:1995df} 
  G.~Jungman, M.~Kamionkowski and K.~Griest,
  Phys.\ Rept.\  {\bf 267}, 195 (1996)
  doi:10.1016/0370-1573(95)00058-5
  [hep-ph/9506380].
  
\bibitem{Bergstrom:2000pn} 
  L.~Bergstr�m,
  Rept.\ Prog.\ Phys.\  {\bf 63}, 793 (2000)
  doi:10.1088/0034-4885/63/5/2r3
  [hep-ph/0002126].
  
\bibitem{Bertone:2004pz} 
  G.~Bertone, D.~Hooper and J.~Silk,
  Phys.\ Rept.\  {\bf 405}, 279 (2005)
  doi:10.1016/j.physrep.2004.08.031
  [hep-ph/0404175].

\bibitem{Feng:2010gw} 
  J.~L.~Feng,
  Ann.\ Rev.\ Astron.\ Astrophys.\  {\bf 48}, 495 (2010)
  doi:10.1146/annurev-astro-082708-101659
  [arXiv:1003.0904 [astro-ph.CO]].


\bibitem{Silveira:1985rk} 
  V.~Silveira and A.~Zee,
  Phys.\ Lett.\  {\bf 161B}, 136 (1985).
  doi:10.1016/0370-2693(85)90624-0
  
\bibitem{Patt:2006fw} 
  B.~Patt and F.~Wilczek,
  hep-ph/0605188.

\bibitem{Galison:1983pa} 
  P.~Galison and A.~Manohar,
  Phys.\ Lett.\  {\bf 136B}, 279 (1984).
  doi:10.1016/0370-2693(84)91161-4
  
\bibitem{Holdom:1985ag} 
  B.~Holdom,
  Phys.\ Lett.\  {\bf 166B}, 196 (1986).
  doi:10.1016/0370-2693(86)91377-8
  
\bibitem{seesaw}
P. Minkowski, Phys. Lett. {\bf B67}, 421 (1977);
T. Yanagida, in {\it Proc. of the Workshop on Grand Unified Theory and Baryon Number of the Universe}, KEK, Japan, 1979;
M. Gell-Mann, P. Ramond  and R. Slansky in {\it Sanibel Symposium}, February 1979, CALT-68-709 [{\tt retroprint arXiv:hep-ph/9809459}],
 and in {\it Supergravity}, eds. D. Freedman {\it et al}. (North Holland, Amsterdam, 1979);
 S.~L. Glashow in {\it Quarks and Leptons, Cargese,} eds. M. Levy {\it et al.} (Plenum, 1980, New York), p. 707;
 R. N. Mohapatra and G. Senjanovic,  Phys. Rev. Lett. {\bf 44}, 912 (1980);
  J.~Schechter and J.~W.~F.~Valle,  Phys.\ Rev.\ D {\bf 22}, 2227 (1980).


\bibitem{McDonald:1993ex} 
  J.~McDonald,
  Phys.\ Rev.\ D {\bf 50}, 3637 (1994)
  doi:10.1103/PhysRevD.50.3637
  [hep-ph/0702143 [HEP-PH]].
  
\bibitem{Burgess:2000yq} 
  C.~P.~Burgess, M.~Pospelov and T.~ter Veldhuis,
  Nucl.\ Phys.\ B {\bf 619}, 709 (2001)
  doi:10.1016/S0550-3213(01)00513-2
  [hep-ph/0011335].
  
\bibitem{Assamagan:2016azc} 
  K.~Assamagan {\it et al.},
  arXiv:1604.05324 [hep-ph].
  
\bibitem{Pospelov:2007mp} 
  M.~Pospelov, A.~Ritz and M.~B.~Voloshin,
  Phys.\ Lett.\ B {\bf 662}, 53 (2008)
  doi:10.1016/j.physletb.2008.02.052
  [arXiv:0711.4866 [hep-ph]].

\bibitem{ArkaniHamed:2008qn} 
  N.~Arkani-Hamed, D.~P.~Finkbeiner, T.~R.~Slatyer and N.~Weiner,
  Phys.\ Rev.\ D {\bf 79}, 015014 (2009)
  doi:10.1103/PhysRevD.79.015014
  [arXiv:0810.0713 [hep-ph]].
  
\bibitem{Alexander:2016aln} 
  J.~Alexander {\it et al.},
  arXiv:1608.08632 [hep-ph].





\bibitem{Tang:2016sib} 
  Y.~L.~Tang and S.~h.~Zhu,
  arXiv:1609.07841 [hep-ph].
  
  
\bibitem{Tang:2015coo} 
  Y.~L.~Tang and S.~h.~Zhu,
  JHEP {\bf 1603}, 043 (2016)
  doi:10.1007/JHEP03(2016)043
  [arXiv:1512.02899 [hep-ph]].
  

  
\bibitem{TheFermi-LAT:2015kwa} 
  M.~Ajello {\it et al.} [Fermi-LAT Collaboration],
  Astrophys.\ J.\  {\bf 819}, no. 1, 44 (2016)
  doi:10.3847/0004-637X/819/1/44
  [arXiv:1511.02938 [astro-ph.HE]].

\bibitem{Goodenough:2009gk} 
  L.~Goodenough and D.~Hooper,
  arXiv:0910.2998 [hep-ph].
  
\bibitem{Hooper:2010mq} 
  D.~Hooper and L.~Goodenough,
  Phys.\ Lett.\ B {\bf 697}, 412 (2011)
  doi:10.1016/j.physletb.2011.02.029
  [arXiv:1010.2752 [hep-ph]].
  
\bibitem{Daylan:2014rsa} 
  T.~Daylan, D.~P.~Finkbeiner, D.~Hooper, T.~Linden, S.~K.~N.~Portillo, N.~L.~Rodd and T.~R.~Slatyer,
  Phys.\ Dark Univ.\  {\bf 12}, 1 (2016)
  doi:10.1016/j.dark.2015.12.005
  [arXiv:1402.6703 [astro-ph.HE]].
  
\bibitem{Calore:2014xka} 
  F.~Calore, I.~Cholis and C.~Weniger,
  JCAP {\bf 1503}, 038 (2015)
  doi:10.1088/1475-7516/2015/03/038
  [arXiv:1409.0042 [astro-ph.CO]].
  


  

\bibitem{Campos:2017odj} 
  M.~D.~Campos, F.~S.~Queiroz, C.~E.~Yaguna and C.~Weniger,
  arXiv:1702.06145 [hep-ph].
  
 
  
\bibitem{Falkowski:2009yz} 
  A.~Falkowski, J.~Juknevich and J.~Shelton,
  arXiv:0908.1790 [hep-ph].
  
\bibitem{Kang:2010ha} 
  Z.~Kang and T.~Li,
  JHEP {\bf 1102}, 035 (2011)
  doi:10.1007/JHEP02(2011)035
  [arXiv:1008.1621 [hep-ph]].
  
\bibitem{Falkowski:2011xh} 
  A.~Falkowski, J.~T.~Ruderman and T.~Volansky,
  JHEP {\bf 1105}, 106 (2011)
  doi:10.1007/JHEP05(2011)106
  [arXiv:1101.4936 [hep-ph]].
  
\bibitem{Cherry:2014xra} 
  J.~F.~Cherry, A.~Friedland and I.~M.~Shoemaker,
  arXiv:1411.1071 [hep-ph].
  
\bibitem{Macias:2015cna} 
  V.~Gonzalez Macias and J.~Wudka,
  JHEP {\bf 1507}, 161 (2015)
  doi:10.1007/JHEP07(2015)161
  [arXiv:1506.03825 [hep-ph]].

\bibitem{Gonzalez-Macias:2016vxy} 
  V.~Gonz�lez-Mac�as, J.~I.~Illana and J.~Wudka,
  JHEP {\bf 1605}, 171 (2016)
  doi:10.1007/JHEP05(2016)171
  [arXiv:1601.05051 [hep-ph]].

\bibitem{Escudero:2016tzx} 
  M.~Escudero, N.~Rius and V.~Sanz,
  arXiv:1606.01258 [hep-ph].
  
\bibitem{Escudero:2016ksa} 
  M.~Escudero, N.~Rius and V.~Sanz,
  arXiv:1607.02373 [hep-ph].
  
\bibitem{Allahverdi:2016fvl} 
  R.~Allahverdi, Y.~Gao, B.~Knockel and S.~Shalgar,
  arXiv:1612.03110 [hep-ph].
 
\bibitem{Bertoni:2014mva} 
  B.~Bertoni, S.~Ipek, D.~McKeen and A.~E.~Nelson,
  JHEP {\bf 1504}, 170 (2015)
  doi:10.1007/JHEP04(2015)170
  [arXiv:1412.3113 [hep-ph]].

\bibitem{Ibarra:2016fco} 
  A.~Ibarra, S.~Lopez-Gehler, E.~Molinaro and M.~Pato,
  Phys.\ Rev.\ D {\bf 94}, no. 10, 103003 (2016)
  doi:10.1103/PhysRevD.94.103003
  [arXiv:1604.01899 [hep-ph]].
  
  

\bibitem{Steigman:2012nb} 
  G.~Steigman, B.~Dasgupta and J.~F.~Beacom,
  Phys.\ Rev.\ D {\bf 86}, 023506 (2012)
  doi:10.1103/PhysRevD.86.023506
  [arXiv:1204.3622 [hep-ph]].



\bibitem{Griest:1989wd} 
  K.~Griest and M.~Kamionkowski,
  Phys.\ Rev.\ Lett.\  {\bf 64}, 615 (1990).
  doi:10.1103/PhysRevLett.64.615

 
\bibitem{Boyarsky:2009ix} 
  A.~Boyarsky, O.~Ruchayskiy and M.~Shaposhnikov,
  Ann.\ Rev.\ Nucl.\ Part.\ Sci.\  {\bf 59}, 191 (2009)
  doi:10.1146/annurev.nucl.010909.083654
  [arXiv:0901.0011 [hep-ph]].
  
\bibitem{Ruchayskiy:2012si} 
  O.~Ruchayskiy and A.~Ivashko,
  JCAP {\bf 1210}, 014 (2012)
  doi:10.1088/1475-7516/2012/10/014
  [arXiv:1202.2841 [hep-ph]].
 

\bibitem{Bandyopadhyay:2011qm} 
  P.~Bandyopadhyay, E.~J.~Chun and J.~C.~Park,
  JHEP {\bf 1106}, 129 (2011)
  doi:10.1007/JHEP06(2011)129
  [arXiv:1105.1652 [hep-ph]].



  
  
\bibitem{Alwall:2014hca} 
  J.~Alwall {\it et al.},
  JHEP {\bf 1407}, 079 (2014)
  doi:10.1007/JHEP07(2014)079
  [arXiv:1405.0301 [hep-ph]].

\bibitem{Alva:2014gxa} 
  D.~Alva, T.~Han and R.~Ruiz,
  JHEP {\bf 1502}, 072 (2015)
  doi:10.1007/JHEP02(2015)072
  [arXiv:1411.7305 [hep-ph]].
  
\bibitem{Degrande:2016aje} 
  C.~Degrande, O.~Mattelaer, R.~Ruiz and J.~Turner,
  Phys.\ Rev.\ D {\bf 94}, no. 5, 053002 (2016)
  doi:10.1103/PhysRevD.94.053002
  [arXiv:1602.06957 [hep-ph]].


\bibitem{Sjostrand:2007gs} 
  T.~Sjostrand, S.~Mrenna and P.~Z.~Skands,
  Comput.\ Phys.\ Commun.\  {\bf 178}, 852 (2008)
  doi:10.1016/j.cpc.2008.01.036
  [arXiv:0710.3820 [hep-ph]].




   
\bibitem{Mardon:2009rc} 
  J.~Mardon, Y.~Nomura, D.~Stolarski and J.~Thaler,
  JCAP {\bf 0905}, 016 (2009)
  doi:10.1088/1475-7516/2009/05/016
  [arXiv:0901.2926 [hep-ph]].


\bibitem{Agrawal:2014oha} 
  P.~Agrawal, B.~Batell, P.~J.~Fox and R.~Harnik,
  JCAP {\bf 1505}, 011 (2015)
  doi:10.1088/1475-7516/2015/05/011
  [arXiv:1411.2592 [hep-ph]].


\bibitem{Elor:2015tva} 
  G.~Elor, N.~L.~Rodd and T.~R.~Slatyer,
  Phys.\ Rev.\ D {\bf 91}, 103531 (2015)
  doi:10.1103/PhysRevD.91.103531
  [arXiv:1503.01773 [hep-ph]].
  
\bibitem{Elor:2015bho} 
  G.~Elor, N.~L.~Rodd, T.~R.~Slatyer and W.~Xue,
  JCAP {\bf 1606}, no. 06, 024 (2016)
  doi:10.1088/1475-7516/2016/06/024
  [arXiv:1511.08787 [hep-ph]].
  




\bibitem{Hinshaw:2012aka} 
  G.~Hinshaw {\it et al.} [WMAP Collaboration],
  Astrophys.\ J.\ Suppl.\  {\bf 208}, 19 (2013)
  doi:10.1088/0067-0049/208/2/19
  [arXiv:1212.5226 [astro-ph.CO]].

\bibitem{Story:2012wx} 
  K.~T.~Story {\it et al.},
  Astrophys.\ J.\  {\bf 779}, 86 (2013)
  doi:10.1088/0004-637X/779/1/86
  [arXiv:1210.7231 [astro-ph.CO]].

\bibitem{Hou:2012xq} 
  Z.~Hou {\it et al.},
  Astrophys.\ J.\  {\bf 782}, 74 (2014)
  doi:10.1088/0004-637X/782/2/74
  [arXiv:1212.6267 [astro-ph.CO]].

\bibitem{Sievers:2013ica} 
  J.~L.~Sievers {\it et al.} [Atacama Cosmology Telescope Collaboration],
  JCAP {\bf 1310}, 060 (2013)
  doi:10.1088/1475-7516/2013/10/060
  [arXiv:1301.0824 [astro-ph.CO]].
  
\bibitem{Ade:2015xua} 
  P.~A.~R.~Ade {\it et al.} [Planck Collaboration],
  Astron.\ Astrophys.\  {\bf 594}, A13 (2016)
  doi:10.1051/0004-6361/201525830
  [arXiv:1502.01589 [astro-ph.CO]].
 
 
  
\bibitem{Padmanabhan:2005es} 
  N.~Padmanabhan and D.~P.~Finkbeiner,
  Phys.\ Rev.\ D {\bf 72}, 023508 (2005)
  doi:10.1103/PhysRevD.72.023508
  [astro-ph/0503486].
  
\bibitem{Zhang:2006fr} 
  L.~Zhang, X.~L.~Chen, Y.~A.~Lei and Z.~G.~Si,
  Phys.\ Rev.\ D {\bf 74}, 103519 (2006)
  doi:10.1103/PhysRevD.74.103519
  [astro-ph/0603425].
  
\bibitem{Galli:2009zc} 
  S.~Galli, F.~Iocco, G.~Bertone and A.~Melchiorri,
  Phys.\ Rev.\ D {\bf 80}, 023505 (2009)
  doi:10.1103/PhysRevD.80.023505
  [arXiv:0905.0003 [astro-ph.CO]].
  
\bibitem{Slatyer:2009yq} 
  T.~R.~Slatyer, N.~Padmanabhan and D.~P.~Finkbeiner,
  Phys.\ Rev.\ D {\bf 80}, 043526 (2009)
  doi:10.1103/PhysRevD.80.043526
  [arXiv:0906.1197 [astro-ph.CO]].
  
\bibitem{Kanzaki:2009hf} 
  T.~Kanzaki, M.~Kawasaki and K.~Nakayama,
  Prog.\ Theor.\ Phys.\  {\bf 123}, 853 (2010)
  doi:10.1143/PTP.123.853
  [arXiv:0907.3985 [astro-ph.CO]].
  
\bibitem{Hisano:2011dc} 
  J.~Hisano, M.~Kawasaki, K.~Kohri, T.~Moroi, K.~Nakayama and T.~Sekiguchi,
  Phys.\ Rev.\ D {\bf 83}, 123511 (2011)
  doi:10.1103/PhysRevD.83.123511
  [arXiv:1102.4658 [hep-ph]].
  
\bibitem{Hutsi:2011vx} 
  G.~Hutsi, J.~Chluba, A.~Hektor and M.~Raidal,
  Astron.\ Astrophys.\  {\bf 535}, A26 (2011)
  doi:10.1051/0004-6361/201116914
  [arXiv:1103.2766 [astro-ph.CO]].
  
\bibitem{Galli:2011rz} 
  S.~Galli, F.~Iocco, G.~Bertone and A.~Melchiorri,
  Phys.\ Rev.\ D {\bf 84}, 027302 (2011)
  doi:10.1103/PhysRevD.84.027302
  [arXiv:1106.1528 [astro-ph.CO]].
  
\bibitem{Finkbeiner:2011dx} 
  D.~P.~Finkbeiner, S.~Galli, T.~Lin and T.~R.~Slatyer,
  Phys.\ Rev.\ D {\bf 85}, 043522 (2012)
  doi:10.1103/PhysRevD.85.043522
  [arXiv:1109.6322 [astro-ph.CO]].
  
\bibitem{Slatyer:2012yq} 
  T.~R.~Slatyer,
  Phys.\ Rev.\ D {\bf 87}, no. 12, 123513 (2013)
  doi:10.1103/PhysRevD.87.123513
  [arXiv:1211.0283 [astro-ph.CO]].
  
\bibitem{Galli:2013dna} 
  S.~Galli, T.~R.~Slatyer, M.~Valdes and F.~Iocco,
  Phys.\ Rev.\ D {\bf 88}, 063502 (2013)
  doi:10.1103/PhysRevD.88.063502
  [arXiv:1306.0563 [astro-ph.CO]].
  
\bibitem{Lopez-Honorez:2013lcm} 
  L.~Lopez-Honorez, O.~Mena, S.~Palomares-Ruiz and A.~C.~Vincent,
  JCAP {\bf 1307}, 046 (2013)
  doi:10.1088/1475-7516/2013/07/046
  [arXiv:1303.5094 [astro-ph.CO]].
  
\bibitem{Madhavacheril:2013cna} 
  M.~S.~Madhavacheril, N.~Sehgal and T.~R.~Slatyer,
  Phys.\ Rev.\ D {\bf 89}, 103508 (2014)
  doi:10.1103/PhysRevD.89.103508
  [arXiv:1310.3815 [astro-ph.CO]].
  
\bibitem{Slatyer:2015jla} 
  T.~R.~Slatyer,
  Phys.\ Rev.\ D {\bf 93}, no. 2, 023527 (2016)
  doi:10.1103/PhysRevD.93.023527
  [arXiv:1506.03811 [hep-ph]].


\bibitem{Slatyer:2015kla} 
  T.~R.~Slatyer,
  Phys.\ Rev.\ D {\bf 93}, no. 2, 023521 (2016)
  doi:10.1103/PhysRevD.93.023521
  [arXiv:1506.03812 [astro-ph.CO]].




\bibitem{Navarro:1995iw} 
  J.~F.~Navarro, C.~S.~Frenk and S.~D.~M.~White,
  Astrophys.\ J.\  {\bf 462}, 563 (1996)
  doi:10.1086/177173
  [astro-ph/9508025].
  
\bibitem{Navarro:1996gj} 
  J.~F.~Navarro, C.~S.~Frenk and S.~D.~M.~White,
  Astrophys.\ J.\  {\bf 490}, 493 (1997)
  doi:10.1086/304888
  [astro-ph/9611107].

\bibitem{Einasto} 
J. Einasto Trudy. 1965. Inst.Astrofiz.Alma-Ata.,5,87.


\bibitem{Navarro:2008kc} 
  J.~F.~Navarro {\it et al.},
  Mon.\ Not.\ Roy.\ Astron.\ Soc.\  {\bf 402}, 21 (2010)
  doi:10.1111/j.1365-2966.2009.15878.x
  [arXiv:0810.1522 [astro-ph]].

\bibitem{Springel:2008cc} 
  V.~Springel {\it et al.},
  Mon.\ Not.\ Roy.\ Astron.\ Soc.\  {\bf 391}, 1685 (2008)
  doi:10.1111/j.1365-2966.2008.14066.x
  [arXiv:0809.0898 [astro-ph]].

\bibitem{Blumenthal:1985qy} 
  G.~R.~Blumenthal, S.~M.~Faber, R.~Flores and J.~R.~Primack,
  Astrophys.\ J.\  {\bf 301}, 27 (1986).
  doi:10.1086/163867
  
\bibitem{Ryden:1987ska} 
  B.~S.~Ryden and J.~E.~Gunn,
  Astrophys.\ J.\  {\bf 318}, 15 (1987).
  doi:10.1086/165349
  
\bibitem{Gnedin:2004cx} 
  O.~Y.~Gnedin, A.~V.~Kravtsov, A.~A.~Klypin and D.~Nagai,
  Astrophys.\ J.\  {\bf 616}, 16 (2004)
  doi:10.1086/424914
  [astro-ph/0406247].


\bibitem{Gnedin:2011uj} 
  O.~Y.~Gnedin, D.~Ceverino, N.~Y.~Gnedin, A.~A.~Klypin, A.~V.~Kravtsov, R.~Levine, D.~Nagai and G.~Yepes,
  arXiv:1108.5736 [astro-ph.CO].
  
    
    
\bibitem{Governato:2012fa} 
  F.~Governato {\it et al.},
  Mon.\ Not.\ Roy.\ Astron.\ Soc.\  {\bf 422}, 1231 (2012)
  doi:10.1111/j.1365-2966.2012.20696.x
  [arXiv:1202.0554 [astro-ph.CO]].
  
  
  
\bibitem{Iocco:2011jz} 
  F.~Iocco, M.~Pato, G.~Bertone and P.~Jetzer,
  JCAP {\bf 1111}, 029 (2011)
  doi:10.1088/1475-7516/2011/11/029
  [arXiv:1107.5810 [astro-ph.GA]].

  
\bibitem{Lee:2014mza} 
  S.~K.~Lee, M.~Lisanti and B.~R.~Safdi,
  JCAP {\bf 1505}, no. 05, 056 (2015)
  doi:10.1088/1475-7516/2015/05/056
  [arXiv:1412.6099 [astro-ph.CO]].
  
\bibitem{Bartels:2015aea} 
  R.~Bartels, S.~Krishnamurthy and C.~Weniger,
  Phys.\ Rev.\ Lett.\  {\bf 116}, no. 5, 051102 (2016)
  doi:10.1103/PhysRevLett.116.051102
  [arXiv:1506.05104 [astro-ph.HE]].
  
\bibitem{Lee:2015fea} 
  S.~K.~Lee, M.~Lisanti, B.~R.~Safdi, T.~R.~Slatyer and W.~Xue,
  Phys.\ Rev.\ Lett.\  {\bf 116}, no. 5, 051103 (2016)
  doi:10.1103/PhysRevLett.116.051103
  [arXiv:1506.05124 [astro-ph.HE]].
  
\bibitem{McDermott:2015ydv} 
  S.~D.~McDermott, P.~J.~Fox, I.~Cholis and S.~K.~Lee,
  JCAP {\bf 1607}, no. 07, 045 (2016)
  doi:10.1088/1475-7516/2016/07/045
  [arXiv:1512.00012 [astro-ph.HE]].
  
\bibitem{Horiuchi:2016zwu} 
  S.~Horiuchi, M.~Kaplinghat and A.~Kwa,
  JCAP {\bf 1611}, no. 11, 053 (2016)
  doi:10.1088/1475-7516/2016/11/053
  [arXiv:1604.01402 [astro-ph.HE]].


\bibitem{Hooper:2012sr} 
  D.~Hooper, C.~Kelso and F.~S.~Queiroz,
  Astropart.\ Phys.\  {\bf 46}, 55 (2013)
  doi:10.1016/j.astropartphys.2013.04.007
  [arXiv:1209.3015 [astro-ph.HE]].


\bibitem{Ackermann:2015zua} 
  M.~Ackermann {\it et al.} [Fermi-LAT Collaboration],
  Phys.\ Rev.\ Lett.\  {\bf 115}, no. 23, 231301 (2015)
  doi:10.1103/PhysRevLett.115.231301
  [arXiv:1503.02641 [astro-ph.HE]].




\bibitem{Aguilar:2016kjl} 
  M.~Aguilar {\it et al.} [AMS Collaboration],
  Phys.\ Rev.\ Lett.\  {\bf 117}, no. 9, 091103 (2016).
  doi:10.1103/PhysRevLett.117.091103



\bibitem{Boudaud:2014qra} 
  M.~Boudaud, M.~Cirelli, G.~Giesen and P.~Salati,
  JCAP {\bf 1505}, no. 05, 013 (2015)
  doi:10.1088/1475-7516/2015/05/013
  [arXiv:1412.5696 [astro-ph.HE]].

\bibitem{Donato:2005my} 
  F.~Donato,
  Nucl.\ Phys.\ Proc.\ Suppl.\  {\bf 138}, 303 (2005).
  doi:10.1016/j.nuclphysbps.2004.11.068



\bibitem{Cirelli:2010xx} 
  M.~Cirelli {\it et al.},
  JCAP {\bf 1103}, 051 (2011)
  Erratum: [JCAP {\bf 1210}, E01 (2012)]
  doi:10.1088/1475-7516/2012/10/E01, 10.1088/1475-7516/2011/03/051
  [arXiv:1012.4515 [hep-ph]].


\bibitem{Aguilar:2015ooa} 
  M.~Aguilar {\it et al.} [AMS Collaboration],
  Phys.\ Rev.\ Lett.\  {\bf 114}, 171103 (2015).
  doi:10.1103/PhysRevLett.114.171103

\bibitem{Giesen:2015ufa} 
  G.~Giesen, M.~Boudaud, Y.~G�nolini, V.~Poulin, M.~Cirelli, P.~Salati and P.~D.~Serpico,
  JCAP {\bf 1509}, no. 09, 023 (2015)
  doi:10.1088/1475-7516/2015/09/023, 10.1088/1475-7516/2015/9/023
  [arXiv:1504.04276 [astro-ph.HE]].
  
\bibitem{Aguilar:2014mma} 
  M.~Aguilar {\it et al.} [AMS Collaboration],
  Phys.\ Rev.\ Lett.\  {\bf 113}, 121102 (2014).
  doi:10.1103/PhysRevLett.113.121102

\bibitem{Adriani:2013uda} 
  O.~Adriani {\it et al.} [PAMELA Collaboration],
  Phys.\ Rev.\ Lett.\  {\bf 111}, 081102 (2013)
  doi:10.1103/PhysRevLett.111.081102
  [arXiv:1308.0133 [astro-ph.HE]].
  
  
\bibitem{Aartsen:2016zhm} 
  M.~G.~Aartsen {\it et al.} [IceCube Collaboration],
  arXiv:1612.05949 [astro-ph.HE].
  
  
\bibitem{Charles:2016pgz} 
  E.~Charles {\it et al.} [Fermi-LAT Collaboration],
  Phys.\ Rept.\  {\bf 636}, 1 (2016)
  doi:10.1016/j.physrep.2016.05.001
  [arXiv:1605.02016 [astro-ph.HE]].


\bibitem{Abbott:2005bi} 
  T.~Abbott {\it et al.} [DES Collaboration],
  astro-ph/0510346.
  
\bibitem{York:2000gk} 
  D.~G.~York {\it et al.} [SDSS Collaboration],
  Astron.\ J.\  {\bf 120}, 1579 (2000)
  doi:10.1086/301513
  [astro-ph/0006396].


\bibitem{Ivezic:2008fe} 
  Z.~Ivezic {\it et al.} [LSST Collaboration],
  arXiv:0805.2366 [astro-ph].




\bibitem{Abdallah:2016ygi} 
  H.~Abdallah {\it et al.} [HESS Collaboration],
  Phys.\ Rev.\ Lett.\  {\bf 117}, no. 11, 111301 (2016)
  doi:10.1103/PhysRevLett.117.111301
  [arXiv:1607.08142 [astro-ph.HE]].
  
  
\bibitem{Smith:2013tta} 
  A.~W.~Smith {\it et al.},
  arXiv:1304.6367 [astro-ph.HE].
  
\bibitem{Aleksic:2013xea} 
  J.~Aleksi? {\it et al.},
  JCAP {\bf 1402}, 008 (2014)
  doi:10.1088/1475-7516/2014/02/008
  [arXiv:1312.1535 [hep-ph]].
  

\bibitem{Doro:2012xx} 
  M.~Doro {\it et al.} [CTA Consortium],
  Astropart.\ Phys.\  {\bf 43}, 189 (2013)
  doi:10.1016/j.astropartphys.2012.08.002
  [arXiv:1208.5356 [astro-ph.IM]].
  
  
  


\bibitem{Abeysekara:2014ffg} 
  A.~U.~Abeysekara {\it et al.} [HAWC Collaboration],
  Phys.\ Rev.\ D {\bf 90}, no. 12, 122002 (2014)
  doi:10.1103/PhysRevD.90.122002
  [arXiv:1405.1730 [astro-ph.HE]].


\bibitem{Moulin:2013lma} 
  E.~Moulin [CTA Consortium],

\bibitem{Silverwood:2014yza} 
  H.~Silverwood, C.~Weniger, P.~Scott and G.~Bertone,
  JCAP {\bf 1503}, no. 03, 055 (2015)
  doi:10.1088/1475-7516/2015/03/055
  [arXiv:1408.4131 [astro-ph.HE]].


\bibitem{Bernlohr:2012we} 
  K.~Bernlöhr {\it et al.},
  Astropart.\ Phys.\  {\bf 43}, 171 (2013)
  doi:10.1016/j.astropartphys.2012.10.002
  [arXiv:1210.3503 [astro-ph.IM]].



  
\bibitem{Mohapatra:1986aw} 
  R.~N.~Mohapatra,
  Phys.\ Rev.\ Lett.\  {\bf 56}, 561 (1986).
  doi:10.1103/PhysRevLett.56.561
  
\bibitem{Atre:2009rg} 
  A.~Atre, T.~Han, S.~Pascoli and B.~Zhang,
  JHEP {\bf 0905}, 030 (2009)
  doi:10.1088/1126-6708/2009/05/030
  [arXiv:0901.3589 [hep-ph]].
  
\bibitem{Deppisch:2015qwa} 
  F.~F.~Deppisch, P.~S.~Bhupal Dev and A.~Pilaftsis,
  New J.\ Phys.\  {\bf 17}, no. 7, 075019 (2015)
  doi:10.1088/1367-2630/17/7/075019
  [arXiv:1502.06541 [hep-ph]].
  
  
\bibitem{Izaguirre:2015pga} 
  E.~Izaguirre and B.~Shuve,
  Phys.\ Rev.\ D {\bf 91}, no. 9, 093010 (2015)
  doi:10.1103/PhysRevD.91.093010
  [arXiv:1504.02470 [hep-ph]].
  
  
\end{thebibliography}
\end{document}